\documentclass[
twocolumn,
]{ceurart}
\conference{}
\usepackage{listings}
\usepackage{subcaption}
\usepackage{comment}
\usepackage{siunitx}
\begin{document}



\title{JanusDDG: A Thermodynamics-Compliant Model for Sequence-Based Protein Stability via Two-Fronts Multi-Head Attention}
%\title{JanusDDG: A Sequence-Based Two-Fronts Multi-Heat Attention Model for State-of-the-Art Prediction of Mutation-Induced Stability Changes}
%\title{JanusDDG: a novel tool to predict protein stability}

\author[1]{Guido Barducci}[%
orcid=0009-0005-1052-8495,
email=guido.barducci@unito.it,
%url=https://bitbucket.org/brainteaser-health/idpp2024-compbiomedunito/,
]

\address[1]{Computational Biomedicine Unit, Dept. of Medical Sciences, University of Turin, Turin, Italy}

\author[1]{Ivan Rossi}[%
orcid=0000-0002-2077-7496,
email=ivan.rossi@unito.it,
%url=https://bitbucket.org/brainteaser-health/idpp2024-compbiomedunito/,
]

\author[1]{Francesco Codicè}[%
orcid=,
email=francesco.codice@unito.it,
%url=https://bitbucket.org/brainteaser-health/idpp2024-compbiomedunito/,
]

\author[1]{Cesare Rollo}[%
orcid=0000-0001-6093-1454,
email=cesare.rollo@unito.it,
%url=https://bitbucket.org/brainteaser-health/idpp2024-compbiomedunito/,
]

\author[1]{Valeria Repetto}[%
orcid=0000-0001-8678-3189,
email=valeria.repetto@unito.it,
%url=https://bitbucket.org/brainteaser-health/idpp2024-compbiomedunito/,
]
\author[1]{Corrado Pancotti}[%
orcid=0000-0003-2327-1148,
email=corrado.pancotti@unito.it,
%url=https://bitbucket.org/brainteaser-health/idpp2024-compbiomedunito/,
]
\author[1]{Virginia Iannibelli}[%
orcid=0009-0008-5253-4051,
email=virginia.iannibelli@unito.it,
%url=https://bitbucket.org/brainteaser-health/idpp2024-compbiomedunito/,
]

\author[1]{Tiziana Sanavia}[%
orcid=0000-0003-3288-0631,
email=tiziana.sanavia@unito.it ,]

\author[1]{Piero Fariselli}[%
orcid=0000-0003-1811-4762,
email=piero.fariselli@unito.it ,
%url=https://bitbucket.org/brainteaser-health/idpp2024-compbiomedunito/,
]

\begin{abstract}
Understanding how residue variations affect protein stability is crucial for designing functional proteins and deciphering the molecular mechanisms underlying disease-related mutations. Recent advances in protein language models (PLMs) have revolutionized computational protein analysis, enabling, among other things,  more accurate predictions of mutational effects.
In this work, we introduce JanusDDG, a deep learning framework that leverages PLM-derived embeddings and a bidirectional cross-attention transformer architecture to predict $\Delta \Delta G$ of single and multiple-residue mutations while simultaneously being constrained to respect fundamental thermodynamic properties, such as antisymmetry and transitivity. 
Unlike conventional self-attention, JanusDDG computes queries (Q) and values (V) as the difference between wild-type and mutant embeddings, while keys (K) alternate between the two. This cross-interleaved attention mechanism enables the model to capture mutation-induced perturbations while preserving essential contextual information.
Experimental results show that JanusDDG achieves state-of-the-art performance in predicting $\Delta \Delta G$ from sequence alone, matching or exceeding the accuracy of structure-based methods for both single and multiple mutations. \\
\textbf{Code Availability:} \url{https://github.com/compbiomed-unito/JanusDDG}

\end{abstract}


\maketitle

\section{Introduction}
\label{sec:introduction}

Protein stability is a fundamental property that determines a protein’s structure, function, and overall behavior in biological systems. One of the most widely used metrics for evaluating protein stability is the change in Gibbs free energy ($\Delta \Delta G$), which quantifies the difference in stability between a wild-type protein and its mutant counterpart. $\Delta \Delta G$ is calculated by comparing the free energy of unfolding for both proteins, providing insight into whether a mutation stabilizes or destabilizes the structure.

In this study, we adopt the convention that a positive $\Delta \Delta G$ value indicates a stabilizing mutation (i.e., the mutant form is more thermodynamically favorable than the wild type). Conversely, a negative $\Delta \Delta G$ suggests that the mutation is destabilizing, making the protein more prone to unfolding or degradation. Accordingly, the $\Delta \Delta G$ between a wild-type ($w$) protein and a mutant ($m$) of the same protein is defined as: \begin{equation} \Delta \Delta G = \Delta G_{w} - \Delta G_{m}
\end{equation} where $\Delta G_{w}$ and $\Delta G_{m}$ are given by: \begin{equation} \Delta G_{w} = G^u_{w} - G^f_{w}, \quad \Delta G_{m} = G^u_{m} - G^f_{m}. \end{equation} Here, $G^u$ represents the Gibbs free energy of the unfolded state, while $G^f$ corresponds to that of the folded state.

The $\Delta \Delta G$ analysis is useful in several fields, including protein engineering \cite{shi2025recent}, to design more stable or functional proteins for industrial and medical applications \cite{gebauer2020engineered}, in drug discovery \cite{meghwanshi2020enzymes}, where it can guide the development of small molecules that either compensate for destabilizing mutations or exploit structural weaknesses in pathogenic proteins, and in medicine, where it helps to predict the impact of genetic mutations in disorders caused by protein misfolding, such as Alzheimer disease \cite{mehra2019computational}, amyotrophic lateral sclerosis \cite{pancotti2022deep},  and cystic fibrosis  \cite{bahia2021stability}.

In recent years, many studies have been published where the prediction of $\Delta \Delta G$ is based on energy-based force fields\cite{Delgado2025,dehouck2011popmusic,Hernandez2023} , machine learning \cite{Sanavia2020} and deep learning models \cite{Pancotti2021, Benevenuta2023, Umerenkov2023, Yang2023, Jiang2024,Li2024a,Cuturello2024,Dieckhaus2024,Chu2024,Sun2025}, mostly based on protein language models\cite{plm2025,Ruffolo2024, savojardo2025ddgemb}. These methods belong to two main categories: sequence-based models and structure-based models. The former are more convenient to use because they require as input just the amino acid sequence of the protein and its mutations, while the latter also need the spatial structure of the protein.  Structure-based methods have usually shown better performance than sequence-based ones \cite{pucci2018quantification,pancotti2022b,reeves2024zero, rollo2023influence} thus justifying the non-trivial extra requirement of producing a realistic model of the protein structure when the experimental structural information is not already available. The latter can be performed using tools such as AlphaFold\cite{Jumper2021}, RoseTTaFold\cite{Baek2021} and OpenFold\cite{ahdritz2024openfold}, that can produce high-quality structures but that are also computationally expensive.
Furthermore, most of the systems developed so far share the limitation of predicting stability changes just for single or double mutations~\cite{mutateeverything2023,Dieckhaus2025}. Few models to date are capable of predicting mutations involving more than two amino acids~\cite{Delgado2025,montanucci2022ddgun,chen2024,savojardo2025ddgemb}.

In this work, we introduce JanusDDG, a deep learning model that leverages a protein language model~\cite{lin2022language} to extract informative representations from the sequences of wild-type and mutated proteins. By relying solely on sequence information, JanusDDG avoids the need for structural input while still capturing the rich contextual and evolutionary signals encoded in the language model embeddings.
Furthermore, JanusDDG is explicitly designed to be compliant with the physics of protein stability. Through the use of tailored loss functions and architectural constraints, it enforces fundamental thermodynamic properties such as \textit{antisymmetry} (the prediction changes sign when the mutation is reversed) and \textit{transitivity} (mutational effects remain consistent across intermediate states). This ensures that the model's predictions are not only accurate but also physically grounded.

\section{Results}
\label{sec:results}

\begin{figure*}[ht]
    \centering
    \includegraphics[width=.9\textwidth]{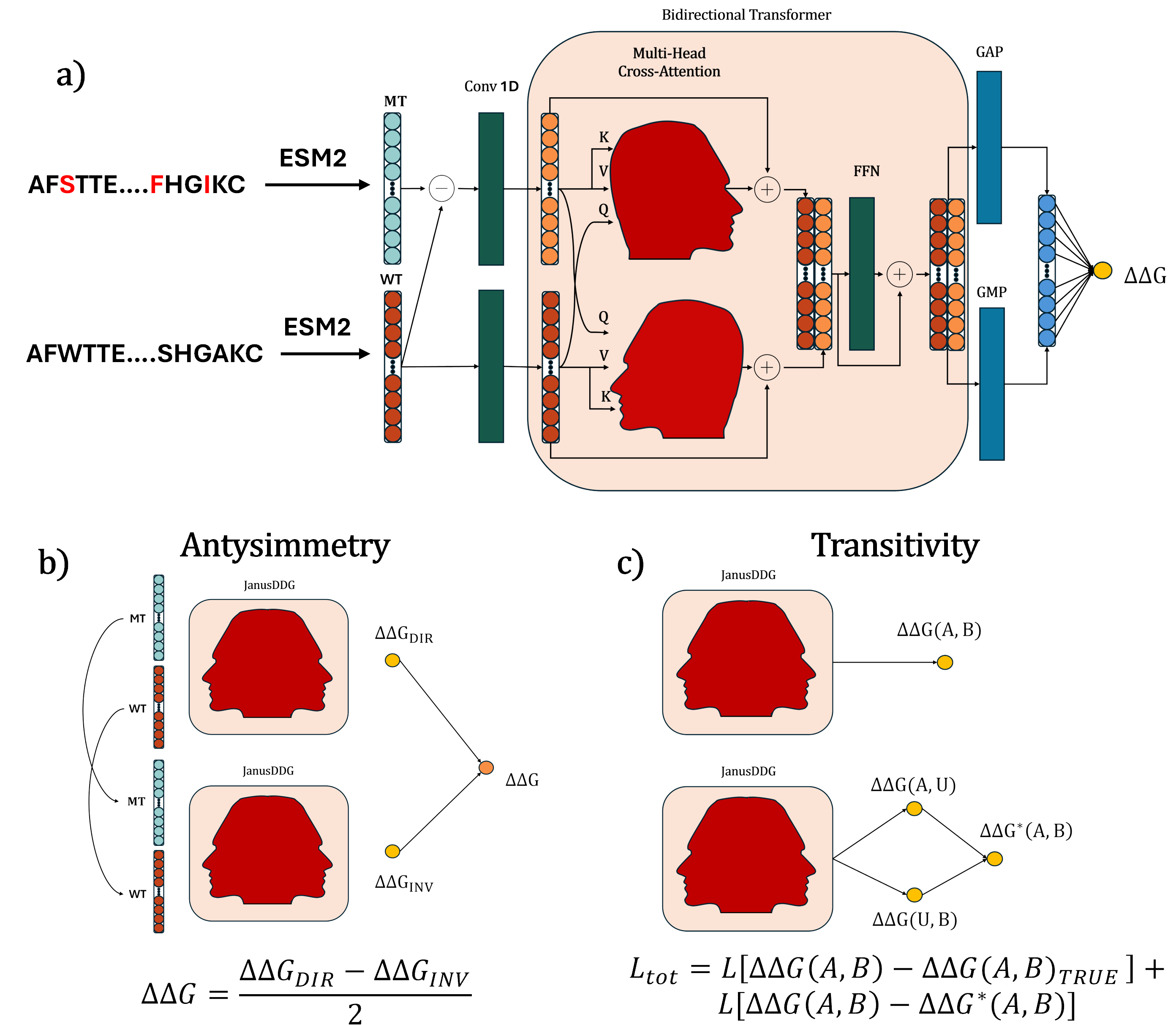} % Specifica il percorso dell'immagine
    \caption[Overwie of JanusDDG.]{\textbf{Overwie JanusDDG.}
    
    \textbf{a) JanusDDG Backbone.}
    The model takes as input wild-type and mutant-type amino acid sequences without requiring 3D structural information, and provides a prediction of $\Delta\Delta G$ by leveraging the power of bidirectional cross-attention. This backbone model enables the prediction of stability changes resulting from single and multi mutations, capturing the underlying patterns of sequence-to-stability relationships effectively. \textbf{b) Antisymmetry.}
    To make the JanusDDG model antisymmetric by design, the base JanusDDG backbone is applied twice with inverted inputs. The resulting predictions are subtracted from each other and then divided by 2. This procedure leverages the antisymmetry as a fundamental property of the model, contributing to a more accurate representation of the relationship between mutations and stability changes. \textbf{c) Transitivity.}
    To enhance the transitivity of the model, fine-tuning is implemented based on the thermodynamic property that links the Gibbs free energy changes ($\Delta\Delta G$) between three mutations (A, B, U). The loss function is formulated such that the model learns the following relation:$\Delta\Delta G(A,B)=\Delta\Delta G^*(A,B)\equiv \Delta\Delta G(A,U)+\Delta\Delta G(U,B)$.
    This property stems from the fact that the Gibbs free energy is a state function, allowing the model to learn transitive relationships between mutations. This approach enables JanusDDG to be more robust and accurate in predicting stability changes in mutated protein sequences.}
    \label{fig:overview} % Etichetta per riferimenti incrociati
\end{figure*}

A model capable of predicting protein stability directly from sequence is of significant importance, as it can be applied more easily in practice, since it does not require the $3D$-structures of the proteins of interest. In a recent study, DDGemb was introduced as a sequence-based predictor capable of estimating $\Delta \Delta G$ for both single and double mutations \cite{savojardo2025ddgemb}. This model leverages the ESM2 language model \cite{lin2022language} to generate protein embeddings, which are then processed by a deep learning architecture based on self-attention mechanisms.

Building on the strengths of DDGemb, we explored new ways to enrich the input representation and improve compliance with known physical principles. While DDGemb relies on the difference between wild-type and mutant embeddings, we retain and integrate more of the original contextual information. Additionally, we aimed to design a model architecture that naturally incorporates thermodynamic properties such as antisymmetry and transitivity (Fig.~\ref{fig:overview}), which are foundational to the Gibbs free energy landscape.

To this end, we developed JanusDDG, a novel sequence-based model depicted in Fig.~\ref{fig:overview}. JanusDDG employs bidirectional cross-attention rather than standard self-attention, allowing it to combine information from both the delta between wild-type and mutant sequences and the full wild-type context itself. This design enables JanusDDG to process the entire protein sequence and make predictions for both single and multiple mutations, expanding its scope of application.

In the following sections, we present the results of JanusDDG on widely used benchmark datasets for protein stability prediction, described in detail in Section~\ref{sec:dataset}.

\subsection{State-Function Property of Gibbs Free Energy}
A reliable $\Delta\Delta G$ predictor should reflect the fundamental properties of Gibbs free energy, which is a state function. In particular, two key mathematical properties must be satisfied: 
\begin{itemize} 
\item \textbf{Antisymmetry}: $\Delta\Delta G(A,B) = -\Delta\Delta G(B,A)$
\item \textbf{Transitivity}: $\Delta\Delta G(A,C) = \Delta\Delta G(A,B) + \Delta\Delta G(B,C)$
\end{itemize}
To encourage JanusDDG to respect these properties, we implemented two dedicated strategies, as detailed in the following sections. 

\subsubsection{Antisymmetry}

The $\Delta\Delta G$ prediction must satisfy the property of antisymmetry, which stems from the fundamental thermodynamic principle that Gibbs free energy is a state function (its change depends only on the initial and final states, not on the path taken). Consequently, if a mutation from amino acid $A$ to amino acid $B$ yields a stability change of $\Delta\Delta G(A,B)$, the reverse mutation ($B \rightarrow A$) should result in the opposite change, such that $\Delta\Delta G(B,A) = -\Delta\Delta G(A,B)$.

Failure to satisfy this property would imply an inconsistency in the underlying free energy landscape, violating thermodynamic constraints and potentially leading to unrealistic predictions. Thus, enforcing antisymmetry in $\Delta\Delta G$ estimates is critical for preserving the physical validity of the model.

To impose antisymmetry by design, we adopt a \textit{siamese} neural network architecture, as illustrated in Fig.~\ref{fig:overview}b. Starting from the trained model, which outputs a directional prediction $\Delta\Delta G_{\text{DIR}}$, we construct a mirrored input by swapping the wild-type and mutant sequences to produce a second prediction, $\Delta\Delta G_{\text{INV}}$. The final antisymmetric prediction is then obtained by averaging the two in opposite directions:
\begin{equation}
    \Delta\Delta G = (\Delta\Delta G_{DIR} - \Delta\Delta G_{INV})/2
    \label{eq:antisymmetry}
\end{equation}

To evaluate the impact of enforcing antisymmetry, we assessed JanusDDG on the S669 test dataset (see Materials and Methods). Prior to applying the antisymmetry constraint, the model already exhibited a strong inverse correlation between direct and reverse predictions, with a Pearson correlation coefficient of:
\begin{equation}
    PCC_{d\text{-}r} = -0.95, \quad \langle \delta \rangle = 0.02
\end{equation}
where $\langle \delta \rangle$ denotes the mean absolute deviation from perfect antisymmetry.
However, after introducing the antisymmetry-enforcing modification, the model satisfies the constraint by design. As expected, this resulted in perfect antisymmetric behavior:
\begin{equation}
    PCC_{d\text{-}r} = -1.00, \quad \langle \delta \rangle = 0.00
\end{equation}

When evaluated on the hard SSym benchmark~\cite{pucci2018quantification}, JanusDDG maintains its antisymmetric performance and achieves optimal scores, whereas most state-of-the-art methods fail to meet this criterion (see Supplementary Table~\ref{tab:ssym}).

\subsubsection{Transitivity}

By using the state-function property of Gibbs free energy, if we know the $\Delta \Delta G(A,B)$ and $\Delta \Delta G(B,C)$, we can find the $\Delta \Delta G(A,C)$ by subtracting these two quantities:
\begin{align}
    \Delta \Delta G(A,C) &= \Delta G(A) - \Delta G(C) \notag \\
    &= \Delta G(A) - \Delta G(B) + \Delta G(B) - \Delta G(C) \notag \\
    &= \Delta \Delta G(A,B) + \Delta \Delta G(B,C)
    \label{eq:transitivity}
\end{align}
A $\Delta\Delta G$ prediction model should therefore satisfy this property.
For this reason, the following fine-tuning, shown in Fig. \ref{fig:overview}c, was performed to encourage the model to respect this property.  
JanusDDG was further trained for additional epochs with the introduction of an extra loss function.  
This loss function was designed to train the model to predict $\Delta\Delta G(A,B)$ between the wild-type protein and the mutant, transitioning through an additional protein state. The predicted value should match $\Delta\Delta G(A,B)$.
The final loss function is therefore given by:  
\begin{equation}
    \begin{aligned}
        L_{TOT} &= L[\Delta\Delta G(A,B)-\Delta\Delta G_{TRUE}] +  \\ 
        &\quad L\{\Delta\Delta G(A,B)- \\
        &\quad [\Delta\Delta G(A,X)+\Delta\Delta G(X,B)]\}
    \end{aligned}
\end{equation}
More technical details are provided in Section \ref{sec:Fine-Tuning}. 

\begin{figure*}[ht]
    \centering
    \includegraphics[width=0.9\textwidth]{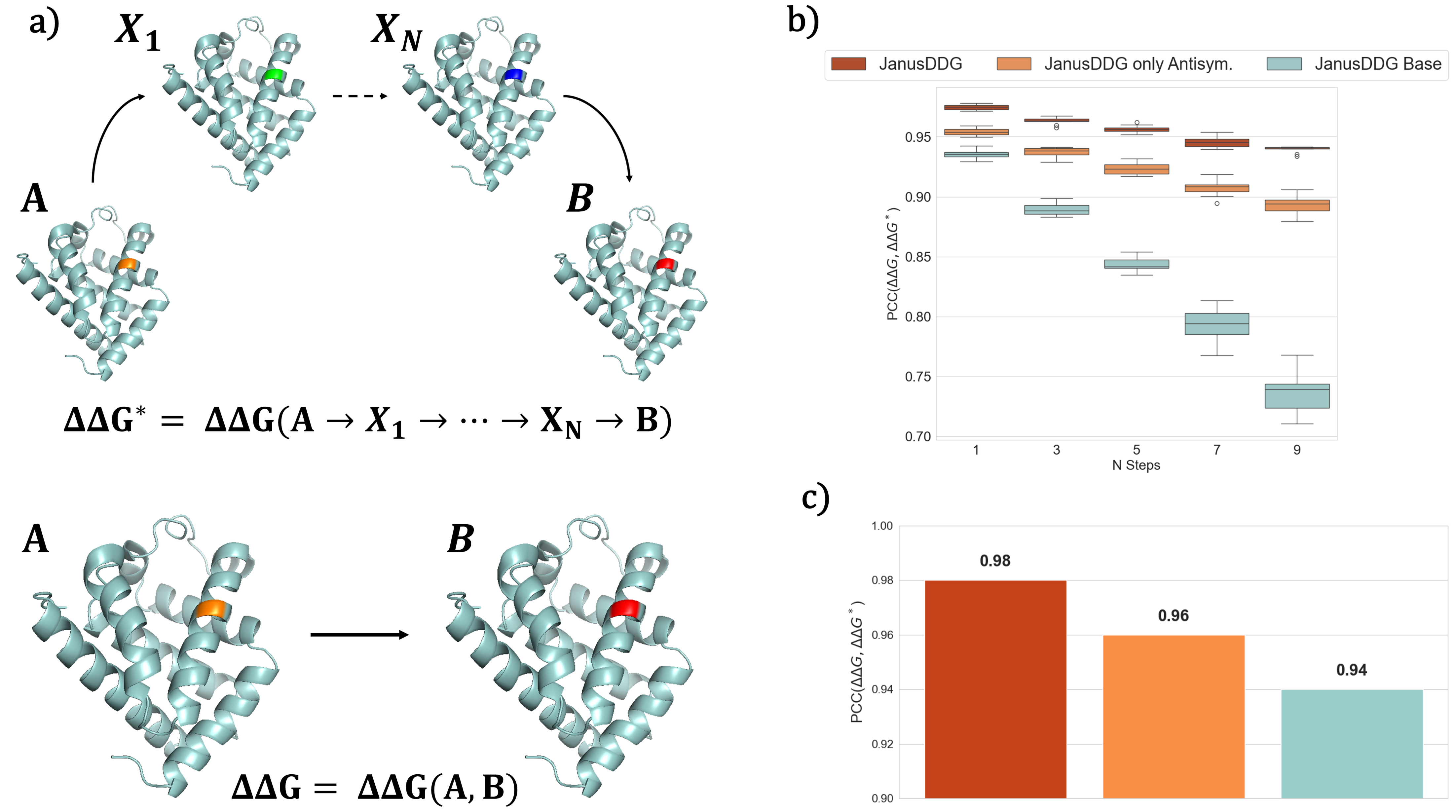} % Specifica il percorso dell'immagine
    \caption{\textbf{Results of the transitivity evaluation of JanusDDG.}  
\textbf{(a)} Illustration of the transitivity property: Since $\Delta\Delta G$ depends only on the initial and final states, $\Delta\Delta G(A,B)$ should be equal to $\Delta\Delta G^*(A,B)$, where the latter is computed by summing the $\Delta\Delta G$ values of multiple intermediate mutations from step 1 to N. The protein figures have been created using PyMOL~\cite{PyMOL}.  
\textbf{(b)} Pearson correlation results between $\Delta\Delta G$ and $\Delta\Delta G^*$, calculated on S669 for different intermediate steps (1, 3, 5, 7, and 9). For each step, the Pearson correlation was computed 10 times for three different models: JanusDDG Base (the model without antisymmetry and fine-tuning), JanusDDG only Antisym. (the model with antisymmetry but without fine-tuning), and JanusDDG (the final model, incorporating both antisymmetry and fine-tuning).  
\textbf{(c)} Transitivity performance, evaluated on the external dataset $\text{S}^{\text{transitive}}$, for all three models.
}
    \label{fig:transitivity_janus_bello} % Etichetta per riferimenti incrociati
\end{figure*}

To assess the extent to which JanusDDG satisfies the transitivity property, we evaluated the model on both the S669 test set (by introducing random intermediate residues) and the $\text{S}^{\text{transitive}}$ dataset, which was specifically developed to evaluate transitivity in $\Delta\Delta G$ prediction~\cite{Samaga2021}. 
In detail, we quantified transitivity by computing the Pearson correlation between the direct prediction $\Delta\Delta G(A,B)$ and the corresponding transitive prediction obtained by inserting a variable number of random amino acids (1, 3, 5, 7, or 9) between residues A and B.
The results of this evaluation are presented in Fig.~\ref{fig:transitivity_janus_bello}. 
Interestingly, the explicit incorporation of antisymmetry into the base model already improves transitivity, suggesting a synergistic relationship between these two fundamental thermodynamic constraints. Subsequent fine-tuning with the transitivity loss further amplifies this effect, bringing the model’s behavior into even closer alignment with the expected transitive properties, as further confirmed in Supplementary~\ref{sec:comparison_3janus}.

\subsection{JanusDDG's Performance in Protein Stability Prediction}
To evaluate the performance of JanusDDG, various datasets were used for both single and double mutations. It is known that using test datasets containing proteins similar to those in the training set increases performance, thus limiting the ability to discover the true performance of the model. In this section, we show the predictions of our model on datasets that do not contain proteins with more than 25\% sequence identity to the training set. The performance on other datasets, which might contain similar proteins, is reported in the Supplementary Information.

\subsubsection{Performance on Single Mutations and Multiple Mutations}

We selected three datasets for evaluation: S669~\cite{pancotti2022b}, S461~\cite{Hernandez2023}, and S96. The first two are among the most commonly used benchmarks for this task, while the third was introduced by \cite{montanucci2022ddgun}. These datasets were chosen because they share less than 25\% sequence similarity with the JanusDDG training set, ensuring an unbiased assessment of generalization performance.

\begin{figure*}[htbp]
    \centering
    \begin{subfigure}{0.5\textwidth}
        \centering
        \includegraphics[width=\linewidth]{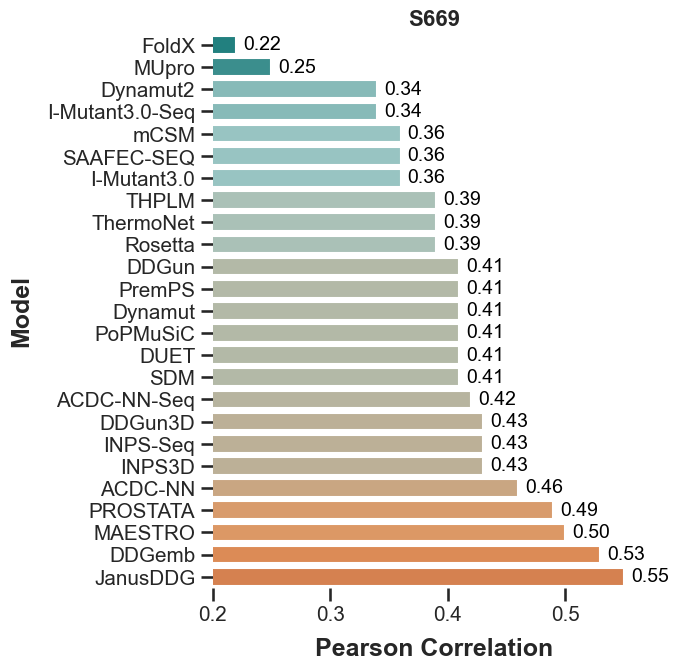}
        \caption{Pearson correlation.}
        \label{fig:pearson_s669_barplot}
    \end{subfigure}%
    \hfill
    \begin{subfigure}{0.5\textwidth}
        \centering
        \includegraphics[width=\linewidth]{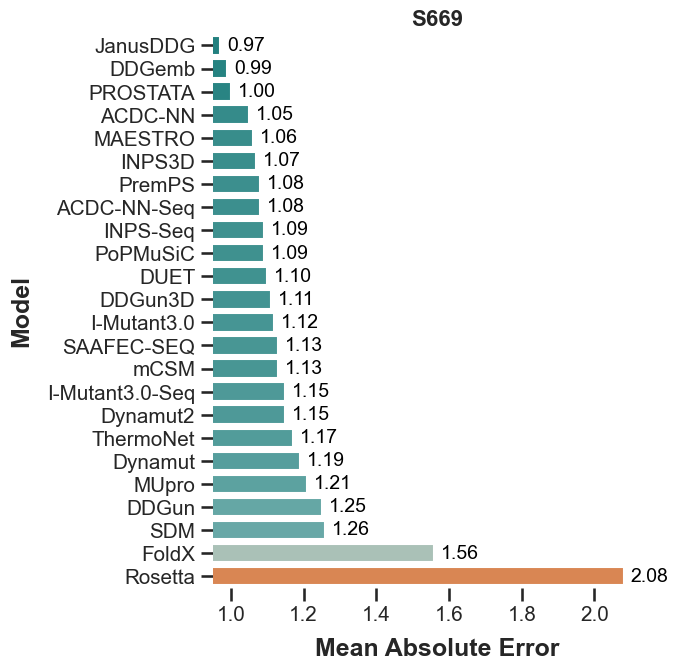}
        \caption{MAE.}
        \label{fig:MAE_s669}
    \end{subfigure}
    \caption{Pearson correlation and MAE on S669 test set. The models' performance data, excluding JanusDDG, were taken from \cite{savojardo2025ddgemb}.}
    \label{fig:s669_pearson_MAE}
\end{figure*}

\begin{figure*}[htbp]
    \centering
    \begin{subfigure}{0.5\textwidth}
        \centering
        \includegraphics[width=\linewidth]{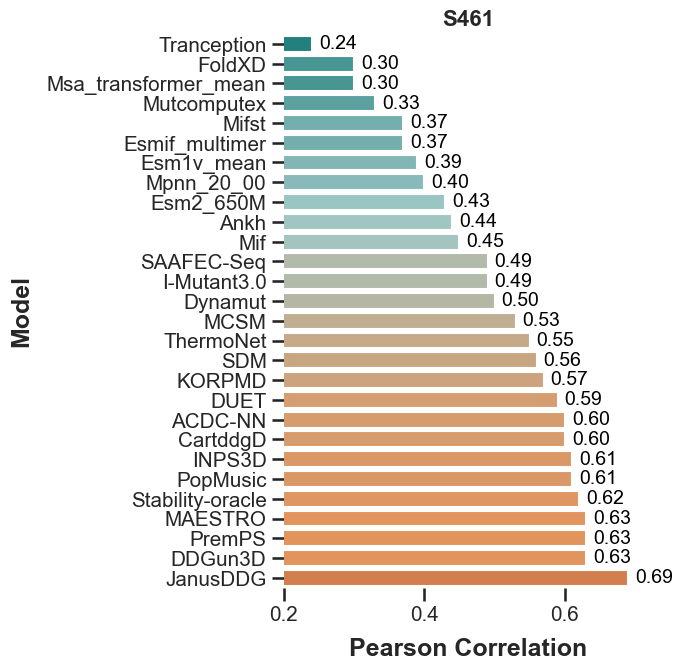}
        \caption{Pearson correlation.}
        \label{fig:pearson_s461}
    \end{subfigure}%
    \hfill
    \begin{subfigure}{0.5\textwidth}
        \centering
        \includegraphics[width=\linewidth]{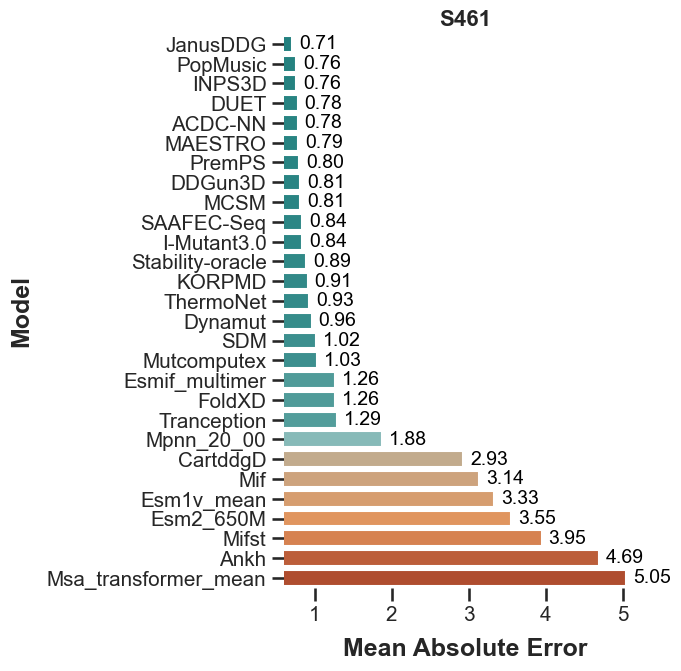}
        \caption{MAE.}
        \label{fig:MAE_s461}
    \end{subfigure}
    \caption{Pearson correlation and MAE on S461 test set. The models' performance data, excluding JanusDDG, were taken from \cite{reeves2024zero}.}
    \label{fig:s461_pearson_MAE}
\end{figure*}

\begin{figure*}[htbp]
    \centering
    \begin{subfigure}{0.5\textwidth}
        \centering
        \includegraphics[width=\linewidth]{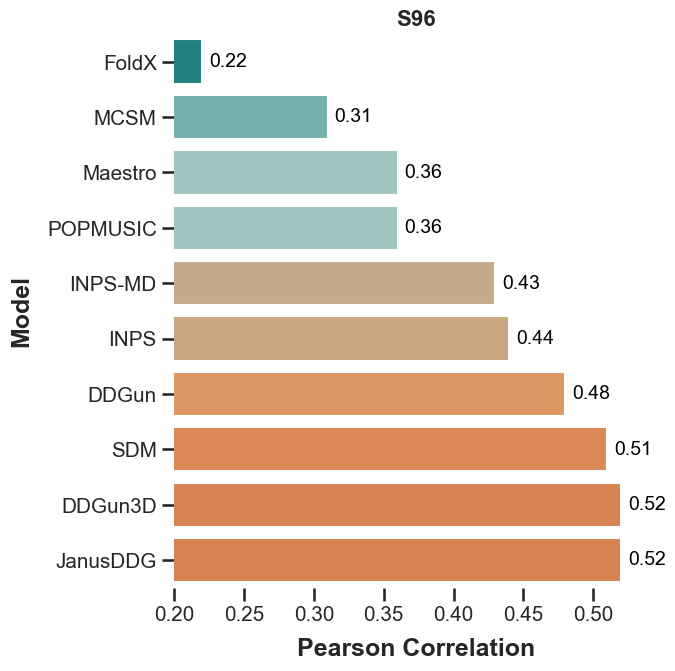}
        \caption{Pearson correlation.}
        \label{fig:S96_pearson}
    \end{subfigure}%
    \hfill
    \begin{subfigure}{0.5\textwidth}
        \centering
        \includegraphics[width=\linewidth]{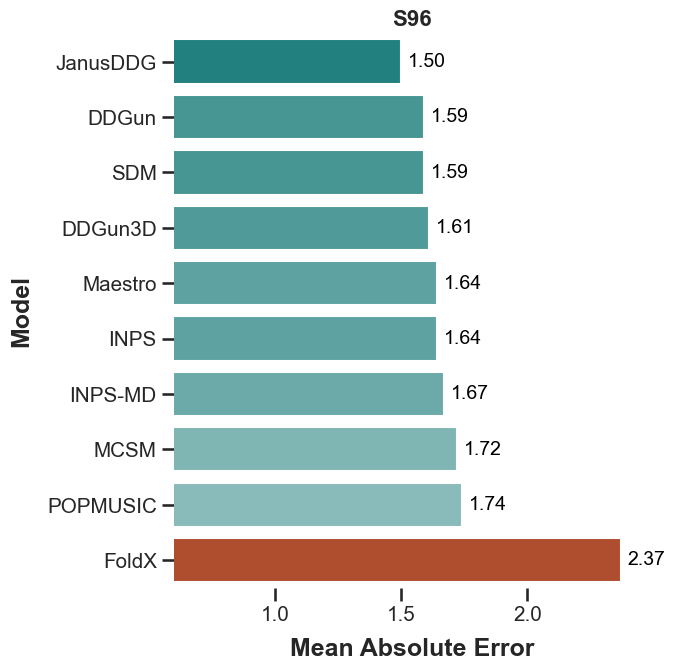}
        \caption{Mean Absolute Error.}
        \label{fig:s96_mae}
    \end{subfigure}
    \caption{Pearson correlation and MAE on S96 test set. The model performance data, excluding JanusDDG, were taken from \cite{montanucci2022ddgun}.}
    \label{fig:s96_pearson_MAE}
\end{figure*}

We compared the performance of JanusDDG using the scores reported in the latest study on these predictions. The results are presented in Fig.~\ref{fig:s669_pearson_MAE} for S669, Fig.~\ref{fig:s461_pearson_MAE} for S461, and Fig.~\ref{fig:s96_pearson_MAE} for S96, where comparisons with other existing models are also provided. Across all three datasets, JanusDDG achieves performance that is comparable to or exceeds that of both existing sequence-based models and several structure-informed predictors, despite relying solely on sequence information. More detailed results are reported in Supplementary Tables~\ref{tab:metrics_single},~\ref{tab:model_performance s461}, and~\ref{tab:performance s96}.

Predicting the stability effects of multiple simultaneous mutations is notably more challenging than single-point mutations, due to potential epistatic interactions. To evaluate JanusDDG in this setting, we used the PTmut-NR dataset~\cite{savojardo2025ddgemb}, which contains proteins with varying numbers of mutations and no close homologs in the training set.

\begin{figure*}[htbp]
    \centering
    \begin{subfigure}{0.5\textwidth}
        \centering
        \includegraphics[width=\linewidth]{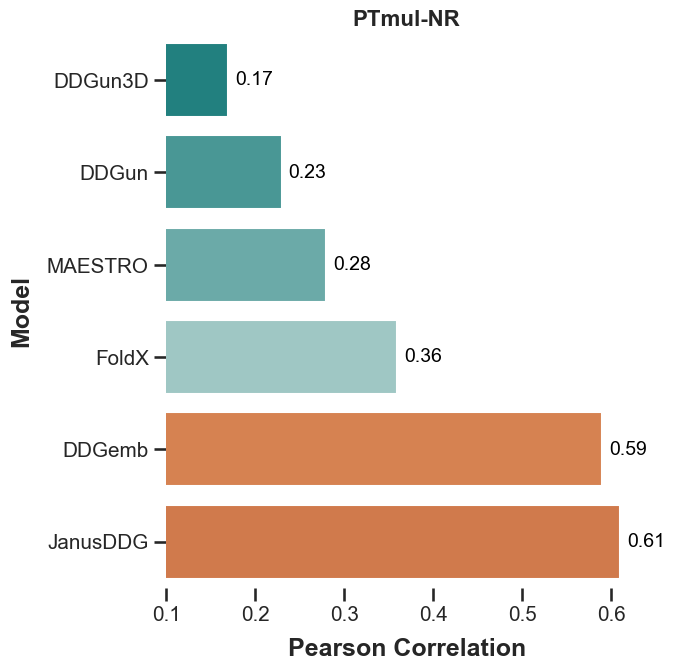}
        \caption{Pearson correlation.}
        \label{fig:ptmul_nr_PEARSON}
    \end{subfigure}%
    \hfill
    \begin{subfigure}{0.5\textwidth}
        \centering
        \includegraphics[width=\linewidth]{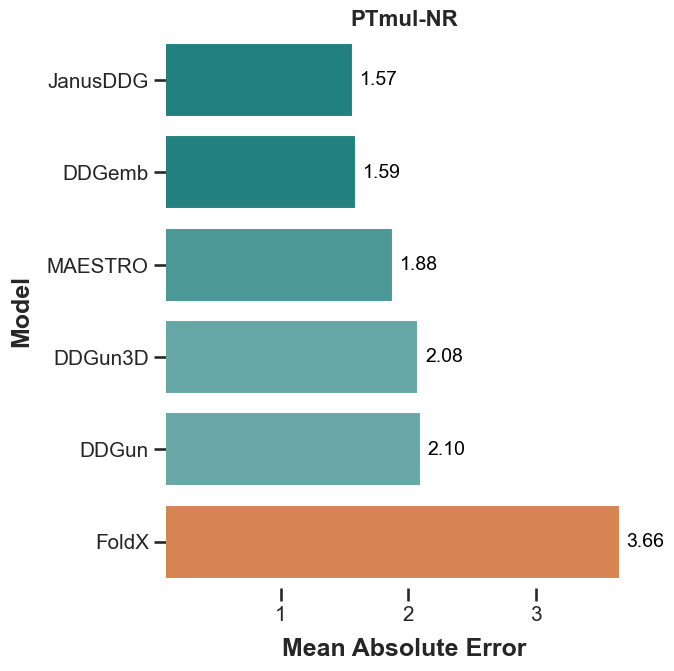}
        \caption{MAE.}
        \label{fig:ptmul_nr_mae}
    \end{subfigure}
    \caption{Pearson correlation and MAE on PTmul-NR test set. The models' performance data, excluding JanusDDG, were taken from \cite{savojardo2025ddgemb}.}
    \label{fig:ptmul_nr_pearson_MAE}
\end{figure*}

The model's performance on this benchmark is reported in Fig.~\ref{fig:ptmul_nr_pearson_MAE} and Table~\ref{tab:performance ptmul_nr}. As with single mutations, JanusDDG outperforms previously published models, demonstrating its ability to generalize to the more complex landscape of multiple-mutation stability prediction.

It is worth noting that, while JanusDDG performs favorably in these benchmarks, relative performance may vary depending on the dataset and experimental conditions, and alternative datasets may yield different model rankings.

\subsubsection{Distance Analysis for Double Mutations}

\begin{figure*}[ht]
    \centering
    \includegraphics[width=0.9\textwidth]{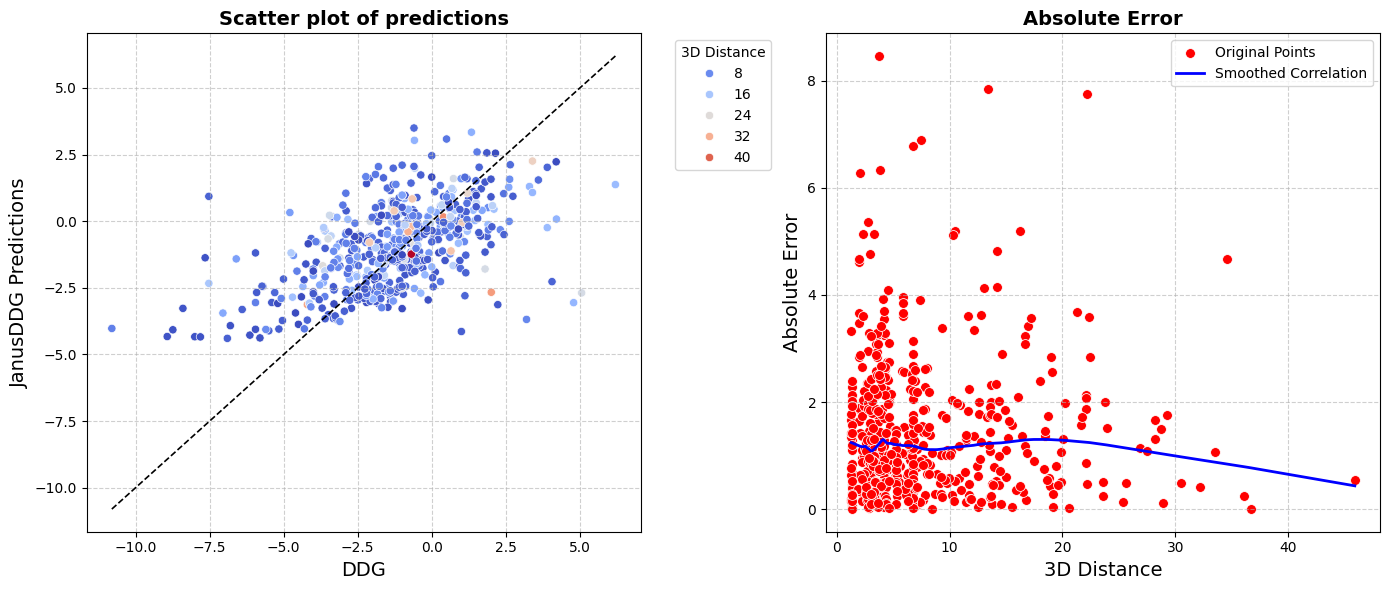} % Specifica il percorso dell'immagine
    \caption{3D Distance Analysis in Double Mutations of Ptmul-D. The left panel shows the correlation between predicted and observed double mutation $\Delta \Delta G$ values, colored by the 3D spatial distance between the mutated residues. The right panel displays the absolute error for each double mutation (red dots) along with a smoothed fitted curve (blue line). }
    \label{fig:3D Distance Analysis} % Etichetta per riferimenti incrociati
\end{figure*}

\begin{figure*}[ht]
    \centering
    \includegraphics[width=1\textwidth]{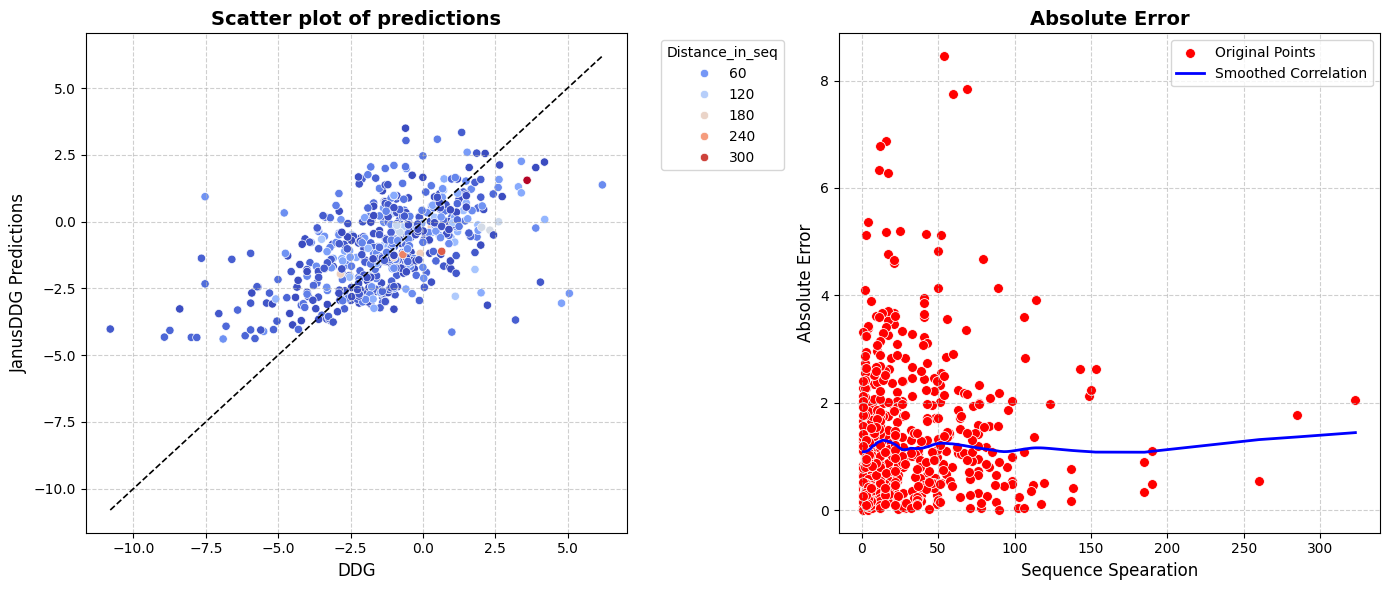} % Specifica il percorso dell'immagine
    \caption{Sequence Separation Analysis in Double Mutations of Ptmul-D. The left panel shows the correlation between predicted and observed double mutation $\Delta \Delta G$ values, colored by the sequence separation between the mutated residues. The right panel displays the absolute error for each double mutation (red dots) along with a smoothed fitted curve (blue line). }
    \label{fig:Sequence Distance Analysis} % Etichetta per riferimenti incrociati
\end{figure*}

It has been observed that deep learning models tend to perform better when the distance between mutated residues is large, as the resulting $\Delta \Delta G$ values exhibit greater additivity~\cite{Dieckhaus2025}. In Figures~\ref{fig:3D Distance Analysis} and~\ref{fig:Sequence Distance Analysis}, we analyze the performance of JanusDDG as a function of the Euclidean 3D distance between mutated residues and their sequence separation, respectively. As an evaluation metric, we use the absolute error between the predicted and experimental $\Delta \Delta G$ values, measured on the PTmul-D dataset. Interestingly, there does not appear to be a significant difference in performance when JanusDDG is evaluated on double mutations that are either close or distant, in terms of both spatial proximity and sequence separation.

\subsubsection{Performance Evaluation of JanusDDG in Stability Classification}

The ability of computational stability predictors to correctly identify mutations that stabilize proteins is an essential prerequisite for accelerating protein engineering workflows. 
%Current evidence, however, reveals that the best available stability predictors have limited accuracy, successfully predicting only approximately 20\% of stabilizing mutations, while the majority of their positive predictions prove to be incorrect. More accurate methods, like free energy perturbation (FEP), based on molecular dynamics, achieve a 50\% success rate, but their computational intensity renders them unsuitable for large-scale applications such as computational deep mutational scans~\cite{broom2020computational}.This discrepancy highlights a strong need to develop a method that combines the reliability of free energy perturbation with significantly greater computational.
%In order to evaluate the performance of JanusDDG concerning this specific capability, we employed the S461 dataset. 
This underscores the strong need for developing a method capable of making such predictions.
In order to evaluate the performance of JanusDDG concerning this specific capability, we employed the S461 dataset. 
%As a preliminary step, we excluded mutations exhibiting an absolute experimental $\Delta\Delta G$ value less than $0.5$ (the average experimental error~\cite{capriotti2008three}). 

\begin{figure*}[ht]
    \centering
    \includegraphics[width=.9\textwidth]{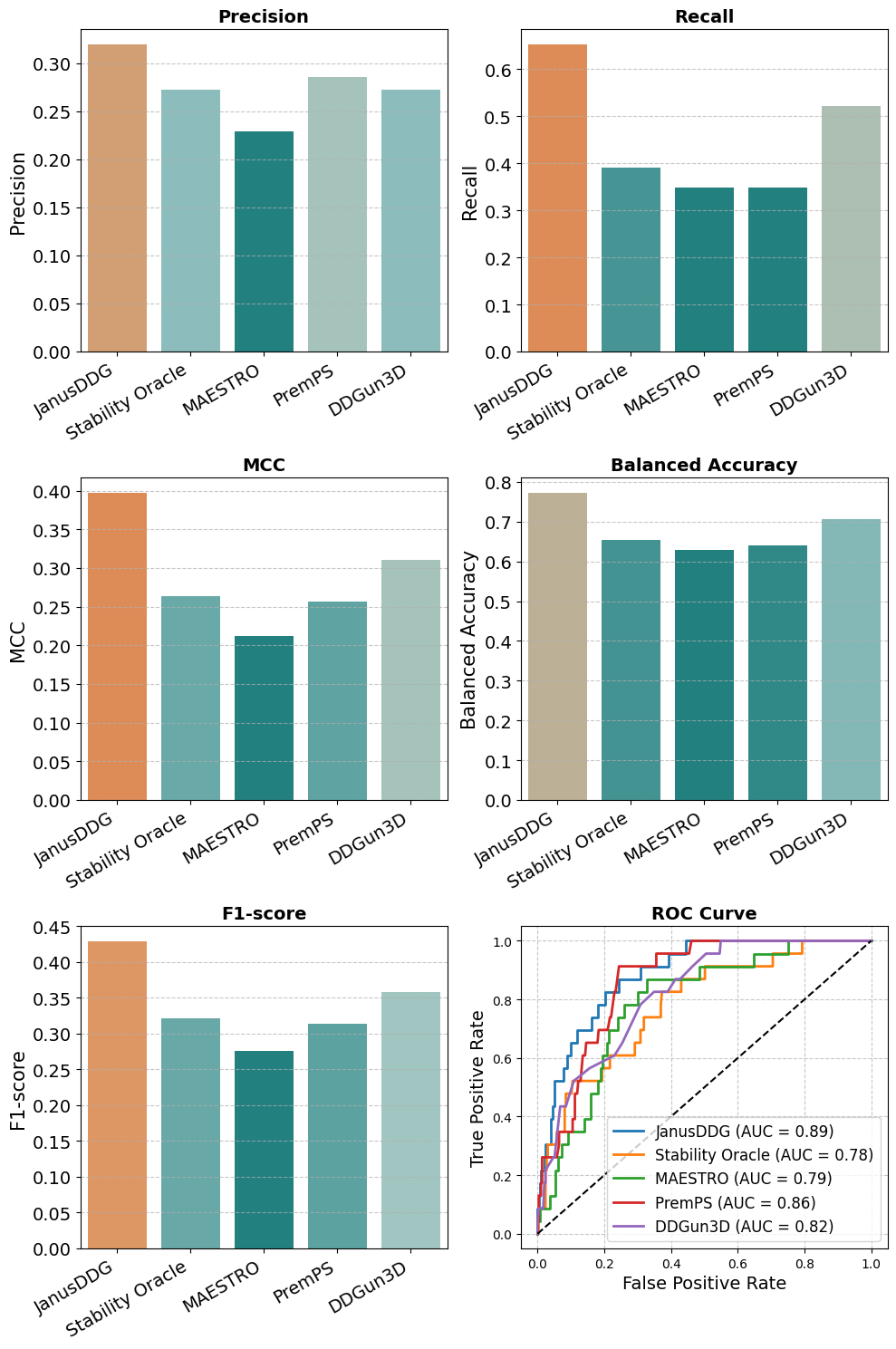} % Specifica il percorso dell'immagine
    \caption{Performance of the top 5 models (based on Pearson correlation on the S461 dataset) in predicting the stability of mutated proteins. The evaluation includes recall, precision, MCC, balanced accuracy, F1 score, and AUC, with ROC curves.}
    \label{fig:classificazione} % Etichetta per riferimenti incrociati
\end{figure*}

Given that the average experimental error of $\Delta\Delta G$ is $0.5\,\text{kcal/mol}$, we define stabilizing proteins as those with $\Delta\Delta G > 0.5\,\text{kcal/mol}$, neutral proteins as those with $-0.5\,\text{kcal/mol} < \Delta\Delta G < 0.5\,\text{kcal/mol}$, and destabilizing proteins as those with $\Delta\Delta G < -0.5\,\text{kcal/mol}$~\cite{capriotti2008three}\cite{Diaz2024}.  
%For mutations with a predicted $\Delta\Delta G > 0.5\,\text{kcal/mol}$ from JanusDDG, the experimental distribution is as follows:  stabilizing mutations: 5 ,  neutral mutations: 4 ,  destabilizing mutations: 5.  Furthermore, among all experimentally stabilizing mutations (295), 217 (74\%) were also predicted as stabilizing. The success rate for predicting stabilizing mutations (83\%) exceeds the typical performance of FEP methods ($\sim50\%$).
As a preliminary step, we excluded neutral mutations, then we selected the top 5 models with the highest Pearson correlation on S461 and subsequently analyzed various performance metrics on stability classification. The result is shown in Figure \ref{fig:classificazione}.As shown, JanusDDG performs well across all metrics on this dataset and tends to outperform the other models, although the precision score indicates that predicting stabilizing variants remains challenging.

%exhibiting an absolute experimental $\Delta\Delta G$ value less than $0.5\,\text{kcal/mol}$ (the average experimental error)

\section{Conclusions and Future Work}
\label{sec:conclusion}

In this work, we introduced JanusDDG, a novel sequence-based deep learning model for predicting protein stability changes upon mutation. JanusDDG effectively captures both contextual and differential information between wild-type and mutant sequences by integrating protein language model embeddings with a bidirectional cross-attention architecture. The model is thermodynamically compliant by design, enforcing antisymmetry and learning transitivity through targeted fine-tuning.

Our benchmarking on datasets with low sequence identity to the training set demonstrates that JanusDDG consistently matches or outperforms both existing sequence-based predictors and structure-informed models. This is particularly noteworthy given that JanusDDG operates solely on sequence data, making it broadly applicable to proteins lacking reliable 3D structural models. Furthermore, JanusDDG generalizes well to the more complex task of predicting the effects of multiple mutations, an area where many current models show limited capabilities.

We acknowledge that model performance may vary across datasets due to differences in experimental protocols, mutation types, and sequence diversity. Future validation across broader mutation spectra and more diverse structural classes may alter the relative performance rankings measured in this paper. Nonetheless, this work illustrates how integrating physical constraints with the representational power of protein language models offers a promising direction for improving both accuracy and interpretability in stability prediction.

\section{Methods}
\label{sec:methodology}
\subsection{Datasets}
\label{sec:dataset}
In this subsection, we show the datasets used for training, validation, and testing our model for both single and multiple mutations.

\subsubsection{Blind Test Datasets}
This section details the blind datasets employed to evaluate JanusDDG and to provide a comparative analysis of its performance against other models. These datasets are specifically composed of proteins exhibiting low sequence similarity (less than 25\%) with the training set, a crucial factor in obtaining a reliable measure of the models' true performance.

\paragraph{S669}

The S669 dataset~\cite{pancotti2022b} is widely recognized as a benchmark for scoring protein stability predictors. The strength of this dataset lies in its construction: it exhibits low sequence redundancy (below 25\% identity) compared to common training datasets like S2648~\cite{dehouck2011popmusic} and VariBench~\cite{nair2013v}, thus facilitating unbiased comparisons. Comprising 1338 single-site mutations, both direct and reverse, across 95 protein chains, S669 provides experimentally determined $\Delta\Delta G$ values, which were retrieved from ThermoMutDB~\cite{xavier2021thermomutdb} and manually verified.

\paragraph{S461}
The S461 dataset~\cite{Hernandez2023} is another widely used dataset to measure the performance on protein stability by predictors. 
This curated dataset addressed some inaccuracies present in the original S669 and excluded mutations potentially involved in natural protein function, such as those at oligomer interfaces. The S461 dataset encompasses experimental structures for 48 wild-type proteins, with a range of 1 to 68 mutations per protein, totaling 461 mutations, each with a single experimental $\Delta\Delta G$ measurement.

\paragraph{S96}

Comprising 96 single-site variants across 14 distinct proteins, the S96 dataset~\cite{montanucci2022ddgun} was assembled using the 2021 version of ProTherm~\cite{nikam2021prothermdb} as its source. Each variant within this dataset was subjected to a rigorous manual checking and correction process, informed by the experimental data presented in the corresponding research articles. Furthermore, to ensure independence from commonly used training data, only those variants whose parent proteins showed less than 25\% sequence identity to any protein in the S2648 and VariBench datasets were included in S96.
In this dataset, when multiple experimental $\Delta\Delta G$ values were reported for the same variant, the average has been taken.

\paragraph{PTmul-NR}

The PTmul-NR~\cite{savojardo2025ddgemb} dataset is a carefully curated subset derived from the original PTmul~\cite{montanucci2019ddgun}, specifically designed to assess model performance in predicting $\Delta\Delta G$ for multi-point mutations, particularly under conditions of low sequence similarity with the S2450 dataset. The original PTmul dataset, which includes 914 multi-point variations across 91 proteins, exhibited substantial sequence overlap with our S2450 training data. As a result of a rigorous removal procedure, the PTmul-NR dataset was created, consisting of 82 multi-point variants across 14 proteins. While this reduction significantly decreased the number of variants, it was crucial for ensuring a more reliable comparison of different methods.

\subsubsection{Training and Validation Datasets}
The datasets underpinning the training and validation of JanusDDG are detailed in this section. The S2450 dataset served as the foundational resource for training JanusDDG, being used both during the initial training phase and in the subsequent fine-tuning phase to enhance its predictive performance. In contrast, the M28 dataset was specifically designated as an independent validation set, used exclusively during the fine-tuning procedure to assess the model's ability to generalize to multi-point mutations.

\paragraph{S2450}
The S2450 dataset, introduced by~\cite{savojardo2025ddgemb}, is a refined version of the established S2648 dataset~\cite{dehouck2011popmusic} and originates from a collection of 2648 single amino acid substitutions across 131 distinct proteins. These mutations have experimentally determined $\Delta\Delta G$ values obtained from the ProTherm database~\cite{bava2004protherm}. While S2648 was created to have low similarity with S669 using sequences from the PDB~\cite{pancotti2022b}, the sequence identity of S2450 was re-evaluated using full-length UniProt sequences. Any protein in S2648 exhibiting more than 25\% sequence identity with a protein in S669 was excluded. This rigorous filtering process resulted in the removal of 18 proteins, encompassing 198 individual mutations, ultimately yielding the S2450 dataset utilized as the training set in this research. To balance this dataset between stabilizing and destabilizing mutations, we used the antisymmetry property: $\Delta \Delta G(B,A) = - \Delta \Delta G(A,B)$ to double the dataset and make it less imbalanced in terms of mutation stability.

\paragraph{M28}
The m28 dataset, a collection of multiple-site variants, was constructed using the 2021 version of ProTherm~\cite{nikam2021prothermdb}. Its selection criteria specifically targeted variants with experimental $\Delta\Delta G$ or $\Delta\Delta G_{H_2O}$ values reported after 2013.
In this dataset, when multiple experimental $\Delta\Delta G$ values were reported for the same variant, the average has been taken.

\subsubsection{Other Datasets}

We evaluated JanusDDG on additional datasets to facilitate a comparative analysis of its performance against other methods documented in the literature. Unlike the strictly blind test sets previously discussed, these datasets may include proteins with sequence similarity exceeding 25\% to our training data. This potential overlap could influence the observed performance, possibly leading to an overestimation of the model’s true capability on unseen data. Consequently, the results obtained from these datasets have been interpreted with caution and carry less weight in our overall assessment compared to the findings from the rigorously blind test sets. For this reason, performance on these datasets is reported only in the Supplementary Section. For details on these datasets, please refer to the cited papers. 
The datasets are as follows:
\begin{itemize}
    \item  \textbf{PTmul-D}, a dataset derived from PTmul, filtered to include only double mutations;
    \item \textbf{K2369}, a dataset containing high sequence identity with S2450, as defined in~\cite{reeves2024zero};
    \item \textbf{Q3421}, another dataset with high sequence identity to S2450, as defined in~\cite{reeves2024zero};
    \item \textbf{Ssym}, a dataset generated to test antisymmetry based on protein structure~\cite{pucci2018quantification};
    \item $\textbf{S}^{\textbf{transitive}}$, a dataset designed to evaluate transitivity of the prediction methods~\cite{Samaga2021}.
\end{itemize}

\subsection{Performance Metrics}
To assess the model's regression performance, we used the following metrics, which are among the most commonly used for this purpose.
\begin{itemize}
    
    \item \textbf{Pearson correlation coefficient} measures the linear relationship between two variables $X$ and $Y$ and is defined as:  
    \[
    r = \frac{\sum_{i=1}^n (X_i - \bar{X})(Y_i - \bar{Y})}{\sqrt{\sum_{i=1}^n (X_i - \bar{X})^2 \sum_{i=1}^n (Y_i - \bar{Y})^2}},
    \]  
    where $\bar{X}$ and $\bar{Y}$ are the means of $X$ and $Y$, respectively. 
    \item The \textbf{Spearman correlation coefficient} evaluates the monotonic relationship between two variables using ranked values, and is given by:  
    \[
    \rho = 1 - \frac{6 \sum_{i=1}^n d_i^2}{n(n^2 - 1)},
    \]  
    where $d_i$ is the difference between the ranks of the $i$-th pair of values and $n$ is the number of data points.
    \item \textbf{Root Mean Square Error (RMSE)} quantifies the average squared difference between predicted values $\hat{y}_i$ and observed values $y_i$ as:  
    \[
    \text{RMSE} = \sqrt{\frac{1}{n} \sum_{i=1}^n (\hat{y}_i - y_i)^2}.
    \]  
    \item  \textbf{Mean Absolute Error (MAE)} measures the average of absolute differences between predictions and observations:
    \[
    \text{MAE} = \frac{1}{n} \sum_{i=1}^n |\hat{y}_i - y_i|.
    \]  
\end{itemize}
Furthermore, we adopted two additional metrics to assess anti-symmetry properties \cite{pucci2018quantification}.
\begin{itemize}
    \item  \textbf{Pearson correlation coefficient between} $p_{\text{dir}}$ \textbf{and} $p_{\text{rev}}$, referred to as \( \text{PCC}_{d-r} \), and is defined as:  
    \[
    \text{PCC}_{d-r} = \text{PCC}(p_{\text{dir}}, p_{\text{rev}}),
    \]  
    where $p_{\text{dir}}$ and $p_{\text{rev}}$ are the predicted values in the direct and reverse directions, respectively.
    \item \textbf{Anti-symmetry bias} \( \langle \delta \rangle \), which quantifies the average deviation between $p_{\text{dir}}$ and $p_{\text{rev}}$ and is computed as:  
    \[
    \langle \delta \rangle = \frac{\sum_{i=1}^N (p_{i, \text{dir}} + p_{i, \text{rev}})}{N},
    \]  
    where $N$ is the total number of observations. 
\end{itemize}

To evaluate classification performance in identifying stabilizing mutations, we used the following metrics.
\begin{itemize}
    \item \textbf{Recall (or Sensitivity)} measures the proportion of actual positive cases that are correctly identified by the model. It is defined as:
    \[
    \text{Recall} = \frac{\text{TP}}{\text{TP} + \text{FN}}
    \]
    where TP represents true positives and FN false negatives.

    \item \textbf{Precision} quantifies the proportion of positive predictions that are actually correct. It is calculated as:
    \[
    \text{Precision} = \frac{\text{TP}}{\text{TP} + \text{FP}}
    \]
    where FP denotes false positives.

    \item \textbf{Matthews Correlation Coefficient (MCC)} provides a balanced measure of classification performance, even for imbalanced datasets. It considers TP, TN (true negatives), FP, and FN in a single metric:
    \[
    \text{MCC} = \frac{\text{TP} \times \text{TN} - \text{FP} \times \text{FN}}{\sqrt{(\text{TP}+\text{FP}) (\text{TP}+\text{FN}) (\text{TN}+\text{FP}) (\text{TN}+\text{FN})}}
    \]

    \item \textbf{Balanced Accuracy} accounts for class imbalance by averaging the recall of each class:
    \[
    \text{Balanced Accuracy} = \frac{\text{Sensitivity} + \text{Specificity}}{2}
    \]
    where specificity is given by:
    \[
    \text{Specificity} = \frac{\text{TN}}{\text{TN} + \text{FP}}
    \]

    \item \textbf{F1 Score} is the harmonic mean of precision and recall, providing a single metric that balances both aspects:
    \[
    F1 = 2 \times \frac{\text{Precision} \times \text{Recall}}{\text{Precision} + \text{Recall}}
    \]

    \item \textbf{ROC Curve and AUC}. The Receiver Operating Characteristic (ROC) curve plots the true positive rate (sensitivity) against the false positive rate at various threshold settings. The Area Under the Curve (AUC) quantifies the overall performance, with a value of 1 indicating a perfect classifier and 0.5 representing a random model.
\end{itemize}

\subsection{Proteins Embedding}
\label{sec:embedding}
In this subsection, we show some characteristics of the proteins embedding that we used as model input.

The embedding used to describe the proteins was obtained from ESM2 with 650M parameters~\cite{lin2022language}. We analyzed these representations to better understand the differences between the embeddings of the wild-type and mutated proteins.

\begin{figure*}[ht]
    \centering
    \includegraphics[width=0.8\textwidth]{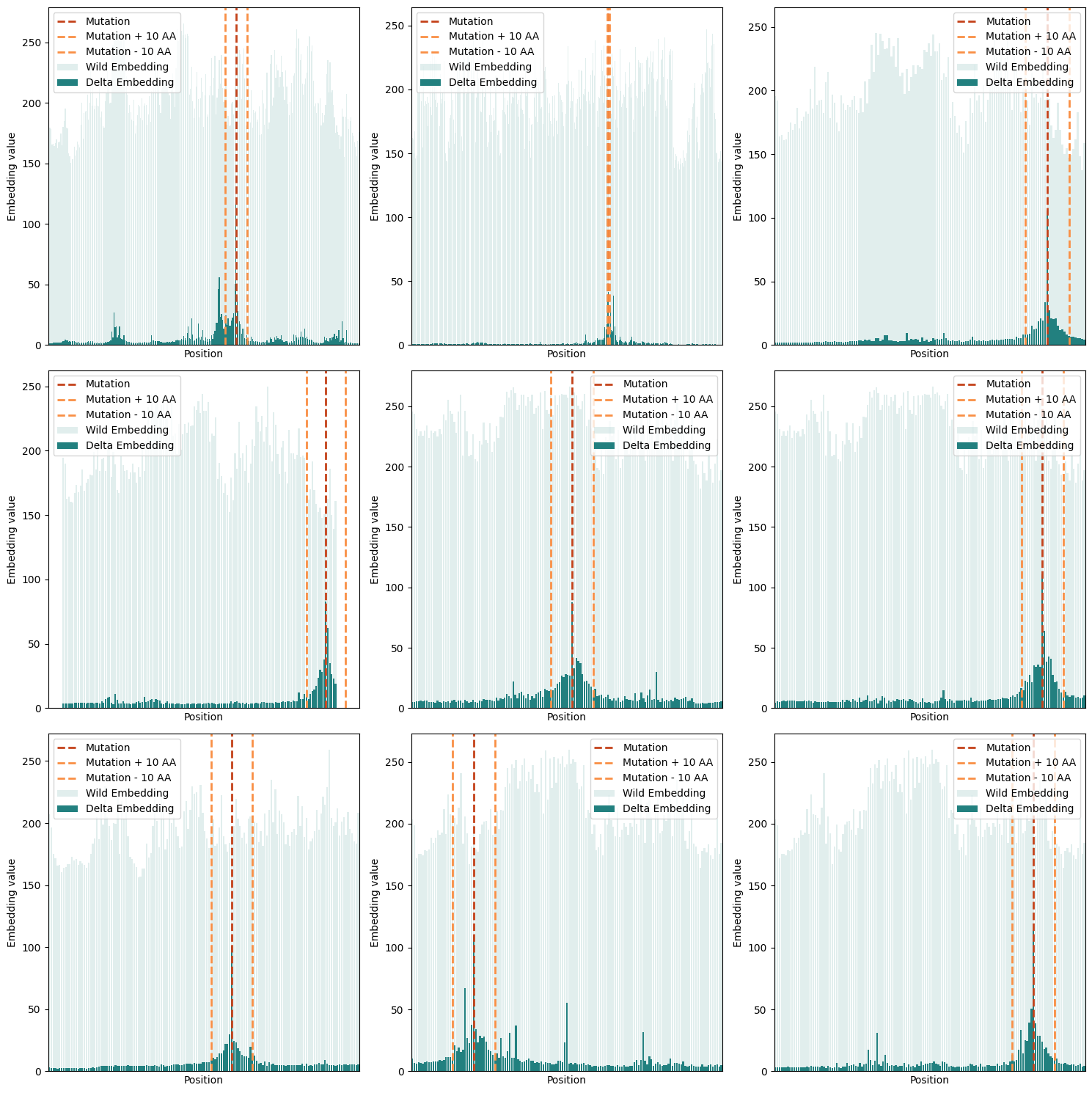} % Specifica il percorso dell'immagine
    \caption{Element-wise absolute difference between the wild-type and mutated protein embeddings (dark blue) with the absolute sum of the elements in the wild-type embedding (light blue).}
    \label{fig:embedding} % Etichetta per riferimenti incrociati
\end{figure*}

The figure~\ref{fig:embedding} presents graphs comparing the element-wise absolute difference between the wild-type and mutated protein embeddings with the absolute sum of the elements in the wild-type embedding. As can be seen, the difference between the mutated and wild-type protein is almost entirely localized around the mutation site.

\begin{figure*}[htbp]
    \centering
    \begin{subfigure}{0.48\textwidth}
        \centering
        \includegraphics[width=\linewidth]{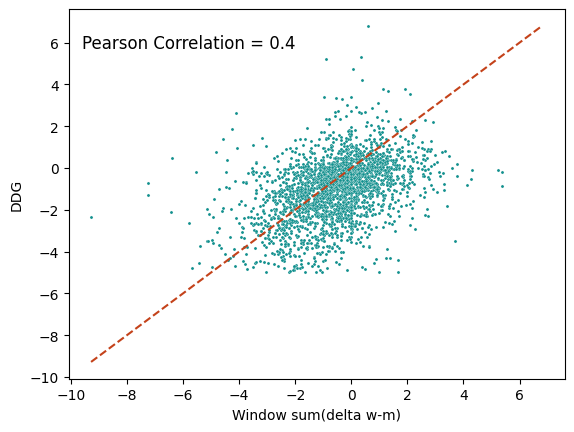}
        \caption{Train}
        \label{fig:sub_train}
    \end{subfigure}%
    \hfill
    \begin{subfigure}{0.48\textwidth}
        \centering
        \includegraphics[width=\linewidth]{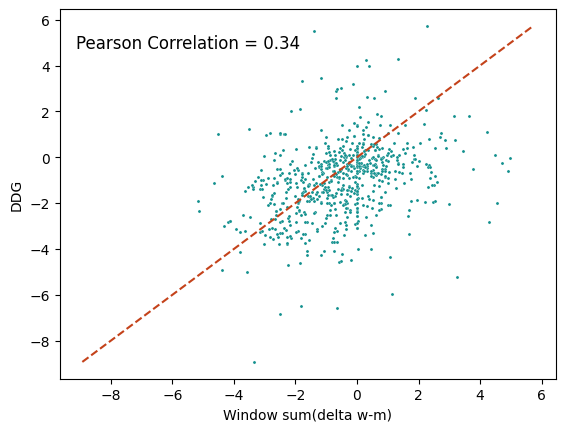}
        \caption{Test}
        \label{fig:sub_test}
    \end{subfigure}
    \caption{Pearson correlation between the difference in ESM2 embeddings of the Wild-type and Mutant-type proteins and the $\Delta\Delta G$.}
    \label{fig:embedding_pearson_corr}
\end{figure*}

To test whether the embedding used is suitable for the prediction task, we tried using the difference between the two embeddings (this time not in absolute value) to predict the $\Delta \Delta G$. This operation is quite naive; therefore, we used only a window around the mutation instead of the entire sequence, as this window contains most of the information about the difference between the wild-type sequence and the mutated sequence. The results are shown in the Figure~\ref{fig:embedding_pearson_corr}. As can be seen, the Pearson correlation achieved on the training and test sets is very high, considering the simplicity of the operation. This suggests that ESM2 can be considered a valid model for extracting the input to be processed by the network.

\subsection{Model Architecture: JanusDDG}

The architecture of our model, shown in Figure \ref{fig:overview}a, is based on the integration of protein language models with the Cross-Attention mechanism. 
The strength of this model is mainly due to two factors: it relies solely on the protein sequence (allowing it to predict the stability of proteins whose structure is unknown) and can be applied regardless of their number of mutations.
This section explains the various building blocks that constitute it.

\subsubsection{Input}
As input to the model, we used the embedding of the two sequences (wild-type and mut-type) obtained from ESM2 650M parameters. These embeddings were also subtracted to derive a third embedding that represents the difference between the two.  
The model takes as input both the wild-type embedding and the difference embedding.

\subsubsection{Conv1D}
Once the input is fed into the model, two 1D convolutions (one for each input) are applied to identify patterns within the sequences while simultaneously reducing their dimensionality. The default Torch parameters were used for the convolution, except for the kernel size, which was set to 20 based on the embedding trend shown in Figure \ref{fig:embedding}.

\subsubsection{Bidirectional Cross Attention Transformer}
The core of the model is the Bidirectional Cross-Attention Transformer. The classic Transformer block consists of the following components: a Multihead Self-Attention block, recurrent connections, and a position-wise FFN. Our proposed model retains all these components except for the first one.

Instead of the Multihead Self-Attention block, we use two Cross-Attention blocks, whose mechanism is explained in the next section. One is applied to the sequence derived from the wild-type sequence embedding, and the other is applied to the sequence derived from the embedding of the difference between the mutated and wild-type sequences.

After applying the two Bidirectional Multihead Cross-Attention blocks, each output is summed with the input of its respective block. Then, the two outputs are concatenated along the feature dimension before being passed into a position-wise FFN.

\paragraph{Bidirectional Cross-Attention}
The standard self-attention mechanism \cite{waswani2017attention} enables each element in a sequence to attend to all others within the same sequence.
In contrast, bidirectional cross-attention extends standard cross-attention by enabling mutual information flow between two input sequences, rather than a one-directional mapping. This mechanism enhances the model’s ability to capture deep interdependencies between two entities.

More specifically, cross-attention is a variant of the attention mechanism where the query $\mathbf{Q}$ comes from one sequence, while the key $\mathbf{K}$ and value $\mathbf{V}$ come from another sequence. Given two input sequences $\mathbf{X} \in \mathbb{R}^{n \times d}$ and $\mathbf{Y} \in \mathbb{R}^{m \times d}$, their corresponding projections are:

\begin{equation}
    \mathbf{Q}_X = \mathbf{X} W_Q, \quad \mathbf{K}_Y = \mathbf{Y} W_K, \quad \mathbf{V}_Y = \mathbf{Y} W_V
\end{equation}

where $W_Q, W_K, W_V \in \mathbb{R}^{d \times d_k}$ are learnable weight matrices.

The attention scores are computed using the scaled dot-product:

\begin{equation}
    \text{Attn}(\mathbf{Q}_X, \mathbf{K}_Y, \mathbf{V}_Y) = \text{softmax} \left( \frac{\mathbf{Q}_X \mathbf{K}_Y^\top}{\sqrt{d_k}} \right) \mathbf{V}_Y
\end{equation}

This allows sequence $\mathbf{X}$ to attend to sequence $\mathbf{Y}$. However, this formulation is inherently asymmetric, meaning that sequence $\mathbf{Y}$ does not simultaneously attend to $\mathbf{X}$.

To capture mutual dependencies, we introduce bidirectional cross-attention, where both sequences $\mathbf{X}$ and $\mathbf{Y}$ attend to each other. This is achieved by computing attention in both directions:

\begin{align}
    \mathbf{H}_X &= \text{softmax} \left( \frac{\mathbf{Q}_X \mathbf{K}_Y^\top}{\sqrt{d_k}} \right) \mathbf{V}_Y \\
    \mathbf{H}_Y &= \text{softmax} \left( \frac{\mathbf{Q}_Y \mathbf{K}_X^\top}{\sqrt{d_k}} \right) \mathbf{V}_X
\end{align}

where:
\begin{itemize}
    \item $\mathbf{H}_X$ represents the updated representation of $\mathbf{X}$ attending to $\mathbf{Y}$.
    \item $\mathbf{H}_Y$ represents the updated representation of $\mathbf{Y}$ attending to $\mathbf{X}$.
\end{itemize}

The final representations are combined through concatenation:

\begin{equation}
    \mathbf{Z} = \textnormal{Cat}(\mathbf{H}_X, \mathbf{H}_Y).
\end{equation}

\subsubsection{Pooling Layers: GAP and GMP}
The output from the previous layer is processed using two different pooling operations: Global Average Pooling and Global Max Pooling. Global Average Pooling computes the average value of each feature map, reducing the spatial dimensions while preserving the overall feature distribution. On the other hand, Global Max Pooling selects the maximum value from each feature map, capturing the most prominent activations. These two operations help distill the most relevant information before passing the representation to the subsequent layers. The outputs of these two layers are concatenated and then passed to the final layer.

\subsubsection{Linear Layer}
To obtain the final $\Delta \Delta G$ value, a linear layer with a single output neuron is applied.

\subsection{Training}
JanusDDG was trained in two distinct phases: a main training phase followed by a fine-tuning phase. Throughout both phases, the model parameters were optimized using the Adam optimizer with Mean Squared Error (MSE) serving as the loss function and a batch size of 6. 

\subsubsection{Main Training Phase}
During the main training phase of JanusDDG, the model was trained on the S2450 dataset augmented with its inverses. The number of training epochs was set to 300, determined via 5-fold cross-validation. In this procedure, we identified the optimal epoch for each fold as the one yielding the maximum Pearson correlation coefficient on the respective validation set. The final count of 300 epochs represents the average of these optimal epoch numbers across the five folds.

\subsubsection{Fine-Tuning Procedure}
\label{sec:Fine-Tuning}

After the main trainig phase we did a fine tuning procedure to try to augmnet perfromance of JanusDDG on two side: Multiple mutations and respect of tranisitivity property.
The transitivity property is one of the two fundamental properties of $\Delta\Delta G$. It states that:

\begin{equation}
    \Delta\Delta G(A,B) = \Delta\Delta G(A,C) + \Delta\Delta G(C,B).
\end{equation}

To enforce transitivity during fine-tuning, we introduce a dedicated two-step loss function (see Fig.~\ref{fig:overview}c) that incorporates a null intermediate state. The second term of the loss, $\text{MSE}(\Delta\Delta G(A,B) - (\Delta\Delta G(A,U) + \Delta\Delta G(U,B)))$, considers the model's prediction by passing through the null state, which serves as a generic thermodynamic reference, enabling the model to evaluate and align mutational effects across multi-step mutation pathways in a physically consistent and residue-agnostic manner.
Furthermore, since a zero vector loses all information about the original amino acids, the model may learn to generalize better to multiple mutations, where the initial and final embeddings differ significantly. As the training dataset, we used S2450 and its inverse, as in the previous training phase. To choose the number of epochs for fine-tuning, we used M28 for validation, tracking the Pearson correlation for each epoch over 30 epochs. The final selected epoch was 28.

\subsection{Hyperparameters Selection}
%For the selection of hyperparameters for our model, we leveraged the fact that the main variations in the ESM2 embedding occur within a 20-amino-acid (AA) window centered on the mutation, leading us to choose a filter size of 20 for the Conv1D layers, as shown in Section~\ref{sec:embedding}. 

Given the long training time of the model and the large number of hyperparameters to be tuned, we opted to adopt the Transformer hyperparameters from DDGemb for our Bidirectional Transformer, as this model employs the same architecture. Specifically, we used the following values: Transformer heads (8), position-wise feed-forward network (FFN) size (512), and the number of filters for the Conv1D layers (128).  
Additionally, we used the same loss function and optimizer as in DDGemb: mean squared error (MSE) loss and the Adam optimizer~\cite{savojardo2025ddgemb}.

To determine the optimal number of training epochs, we conducted a 5-fold cross-validation on the training set. We utilized the five folds defined by \cite{savojardo2025ddgemb}, where MMSeq2~\cite{steinegger2017mmseqs2} was employed to partition the training set into five subsets, each containing proteins with similar sequences. This approach ensured that proteins sharing more than 25\% sequence identity were assigned to the same subset. The final number of training epochs was set to 300, corresponding to the mean of the best epoch for each fold.

%% Forza la stampa di tutte le figure e tabelle prima della bibliografia
\clearpage

%% Define the bibliography file to be used
\bibliography{bibliography}

\begin{thebibliography}{46}
\expandafter\ifx\csname natexlab\endcsname\relax\def\natexlab#1{#1}\fi
\providecommand{\url}[1]{\texttt{#1}}
\providecommand{\href}[2]{#2}
\providecommand{\path}[1]{#1}
\providecommand{\DOIprefix}{doi:}
\providecommand{\ArXivprefix}{arXiv:}
\providecommand{\URLprefix}{URL: }
\providecommand{\Pubmedprefix}{pmid:}
\providecommand{\doi}[1]{\href{http://dx.doi.org/#1}{\path{#1}}}
\providecommand{\Pubmed}[1]{\href{pmid:#1}{\path{#1}}}
\providecommand{\bibinfo}[2]{#2}
\ifx\xfnm\relax \def\xfnm[#1]{\unskip,\space#1}\fi
%Type = Article
\bibitem[{Shi et~al.(2025)Shi, Yuan, Yang, and Sun}]{shi2025recent}
\bibinfo{author}{J.~Shi}, \bibinfo{author}{B.~Yuan}, \bibinfo{author}{H.~Yang},
  \bibinfo{author}{Z.~Sun},
\newblock \bibinfo{title}{Recent advances on protein engineering for improved
  stability},
\newblock \bibinfo{journal}{BioDesign Research}  (\bibinfo{year}{2025})
  \bibinfo{pages}{100005}.
%Type = Article
\bibitem[{Gebauer and Skerra(2020)}]{gebauer2020engineered}
\bibinfo{author}{M.~Gebauer}, \bibinfo{author}{A.~Skerra},
\newblock \bibinfo{title}{Engineered protein scaffolds as next-generation
  therapeutics},
\newblock \bibinfo{journal}{Annual review of pharmacology and toxicology}
  \bibinfo{volume}{60} (\bibinfo{year}{2020}) \bibinfo{pages}{391--415}.
%Type = Article
\bibitem[{Meghwanshi et~al.(2020)Meghwanshi, Kaur, Verma, Dabi, Vashishtha,
  Charan, Purohit, Bhandari, Bhojak, and Kumar}]{meghwanshi2020enzymes}
\bibinfo{author}{G.~K. Meghwanshi}, \bibinfo{author}{N.~Kaur},
  \bibinfo{author}{S.~Verma}, \bibinfo{author}{N.~K. Dabi},
  \bibinfo{author}{A.~Vashishtha}, \bibinfo{author}{P.~Charan},
  \bibinfo{author}{P.~Purohit}, \bibinfo{author}{H.~Bhandari},
  \bibinfo{author}{N.~Bhojak}, \bibinfo{author}{R.~Kumar},
\newblock \bibinfo{title}{Enzymes for pharmaceutical and therapeutic
  applications},
\newblock \bibinfo{journal}{Biotechnology and applied biochemistry}
  \bibinfo{volume}{67} (\bibinfo{year}{2020}) \bibinfo{pages}{586--601}.
%Type = Article
\bibitem[{Mehra and Kepp(2019)}]{mehra2019computational}
\bibinfo{author}{R.~Mehra}, \bibinfo{author}{K.~P. Kepp},
\newblock \bibinfo{title}{Computational analysis of alzheimer-causing mutations
  in amyloid precursor protein and presenilin 1},
\newblock \bibinfo{journal}{Archives of Biochemistry and Biophysics}
  \bibinfo{volume}{678} (\bibinfo{year}{2019}) \bibinfo{pages}{108168}.
%Type = Article
\bibitem[{Pancotti et~al.(2022)Pancotti, Birolo, Rollo, Sanavia, Di~Camillo,
  Manera, Chi{\`o}, and Fariselli}]{pancotti2022deep}
\bibinfo{author}{C.~Pancotti}, \bibinfo{author}{G.~Birolo},
  \bibinfo{author}{C.~Rollo}, \bibinfo{author}{T.~Sanavia},
  \bibinfo{author}{B.~Di~Camillo}, \bibinfo{author}{U.~Manera},
  \bibinfo{author}{A.~Chi{\`o}}, \bibinfo{author}{P.~Fariselli},
\newblock \bibinfo{title}{Deep learning methods to predict amyotrophic lateral
  sclerosis disease progression},
\newblock \bibinfo{journal}{Scientific reports} \bibinfo{volume}{12}
  (\bibinfo{year}{2022}) \bibinfo{pages}{13738}.
%Type = Article
\bibitem[{Bahia et~al.(2021)Bahia, Khazanov, Zhou, Yang, Wang, Hong, Rab,
  Sorscher, Brouillette, Hunt et~al.}]{bahia2021stability}
\bibinfo{author}{M.~S. Bahia}, \bibinfo{author}{N.~Khazanov},
  \bibinfo{author}{Q.~Zhou}, \bibinfo{author}{Z.~Yang},
  \bibinfo{author}{C.~Wang}, \bibinfo{author}{J.~S. Hong},
  \bibinfo{author}{A.~Rab}, \bibinfo{author}{E.~J. Sorscher},
  \bibinfo{author}{C.~G. Brouillette}, \bibinfo{author}{J.~F. Hunt}, et~al.,
\newblock \bibinfo{title}{Stability prediction for mutations in the cytosolic
  domains of cystic fibrosis transmembrane conductance regulator},
\newblock \bibinfo{journal}{Journal of chemical information and modeling}
  \bibinfo{volume}{61} (\bibinfo{year}{2021}) \bibinfo{pages}{1762--1777}.
%Type = Article
\bibitem[{Delgado et~al.(2025)Delgado, Reche, Cianferoni, Orlando, van~der
  Kant, Rousseau, Schymkowitz, and Serrano}]{Delgado2025}
\bibinfo{author}{J.~Delgado}, \bibinfo{author}{R.~Reche},
  \bibinfo{author}{D.~Cianferoni}, \bibinfo{author}{G.~Orlando},
  \bibinfo{author}{R.~van~der Kant}, \bibinfo{author}{F.~Rousseau},
  \bibinfo{author}{J.~Schymkowitz}, \bibinfo{author}{L.~Serrano},
\newblock \bibinfo{title}{{FoldX} force field revisited, an improved version},
\newblock \bibinfo{journal}{Bioinformatics (Oxford, England)}
  \bibinfo{volume}{41} (\bibinfo{year}{2025}) \bibinfo{pages}{btaf064}.
  \DOIprefix\doi{10.1093/bioinformatics/btaf064}.
%Type = Article
\bibitem[{Dehouck et~al.(2011)Dehouck, Kwasigroch, Gilis, and
  Rooman}]{dehouck2011popmusic}
\bibinfo{author}{Y.~Dehouck}, \bibinfo{author}{J.~M. Kwasigroch},
  \bibinfo{author}{D.~Gilis}, \bibinfo{author}{M.~Rooman},
\newblock \bibinfo{title}{Popmusic 2.1: a web server for the estimation of
  protein stability changes upon mutation and sequence optimality},
\newblock \bibinfo{journal}{BMC bioinformatics} \bibinfo{volume}{12}
  (\bibinfo{year}{2011}) \bibinfo{pages}{1--12}.
%Type = Article
\bibitem[{Hernández et~al.(2023)Hernández, Dehouck, Bastolla, López-Blanco,
  and Chacón}]{Hernandez2023}
\bibinfo{author}{I.~M. Hernández}, \bibinfo{author}{Y.~Dehouck},
  \bibinfo{author}{U.~Bastolla}, \bibinfo{author}{J.~R. López-Blanco},
  \bibinfo{author}{P.~Chacón},
\newblock \bibinfo{title}{Predicting protein stability changes upon mutation
  using a simple orientational potential},
\newblock \bibinfo{journal}{Bioinformatics} \bibinfo{volume}{39}
  (\bibinfo{year}{2023}) \bibinfo{pages}{btad011}. \URLprefix
  \url{https://doi.org/10.1093/bioinformatics/btad011}.
  \DOIprefix\doi{10.1093/bioinformatics/btad011}.
%Type = Article
\bibitem[{Sanavia et~al.(2020)Sanavia, Birolo, Montanucci, Turina, Capriotti,
  and Fariselli}]{Sanavia2020}
\bibinfo{author}{T.~Sanavia}, \bibinfo{author}{G.~Birolo},
  \bibinfo{author}{L.~Montanucci}, \bibinfo{author}{P.~Turina},
  \bibinfo{author}{E.~Capriotti}, \bibinfo{author}{P.~Fariselli},
\newblock \bibinfo{title}{Limitations and challenges in protein stability
  prediction upon genome variations: towards future applications in precision
  medicine},
\newblock \bibinfo{journal}{Computational and Structural Biotechnology Journal}
  \bibinfo{volume}{18} (\bibinfo{year}{2020}) \bibinfo{pages}{1968--1979}.
  \URLprefix \url{https://www.csbj.org/article/S2001-0370(20)30343-3/fulltext}.
  \DOIprefix\doi{10.1016/j.csbj.2020.07.011}.
%Type = Article
\bibitem[{Pancotti et~al.(2021)Pancotti, Benevenuta, Repetto, Birolo,
  Capriotti, Sanavia, and Fariselli}]{Pancotti2021}
\bibinfo{author}{C.~Pancotti}, \bibinfo{author}{S.~Benevenuta},
  \bibinfo{author}{V.~Repetto}, \bibinfo{author}{G.~Birolo},
  \bibinfo{author}{E.~Capriotti}, \bibinfo{author}{T.~Sanavia},
  \bibinfo{author}{P.~Fariselli},
\newblock \bibinfo{title}{A {Deep}-{Learning} {Sequence}-{Based} {Method} to
  {Predict} {Protein} {Stability} {Changes} {Upon} {Genetic} {Variations}},
\newblock \bibinfo{journal}{Genes} \bibinfo{volume}{12} (\bibinfo{year}{2021})
  \bibinfo{pages}{911}. \URLprefix
  \url{https://www.mdpi.com/2073-4425/12/6/911}.
  \DOIprefix\doi{10.3390/genes12060911}.
%Type = Article
\bibitem[{Benevenuta et~al.(2023)Benevenuta, Birolo, Sanavia, Capriotti, and
  Fariselli}]{Benevenuta2023}
\bibinfo{author}{S.~Benevenuta}, \bibinfo{author}{G.~Birolo},
  \bibinfo{author}{T.~Sanavia}, \bibinfo{author}{E.~Capriotti},
  \bibinfo{author}{P.~Fariselli},
\newblock \bibinfo{title}{Challenges in predicting stabilizing variations: {An}
  exploration},
\newblock \bibinfo{journal}{Frontiers in Molecular Biosciences}
  \bibinfo{volume}{9} (\bibinfo{year}{2023}) \bibinfo{pages}{1075570}.
  \URLprefix \url{https://www.ncbi.nlm.nih.gov/pmc/articles/PMC9849384/}.
  \DOIprefix\doi{10.3389/fmolb.2022.1075570}.
%Type = Article
\bibitem[{Umerenkov et~al.(2023)Umerenkov, Nikolaev, Shashkova, Strashnov,
  Sindeeva, Shevtsov, Ivanisenko, and Kardymon}]{Umerenkov2023}
\bibinfo{author}{D.~Umerenkov}, \bibinfo{author}{F.~Nikolaev},
  \bibinfo{author}{T.~I. Shashkova}, \bibinfo{author}{P.~V. Strashnov},
  \bibinfo{author}{M.~Sindeeva}, \bibinfo{author}{A.~Shevtsov},
  \bibinfo{author}{N.~V. Ivanisenko}, \bibinfo{author}{O.~L. Kardymon},
\newblock \bibinfo{title}{{PROSTATA}: a framework for protein stability
  assessment using transformers},
\newblock \bibinfo{journal}{Bioinformatics (Oxford, England)}
  \bibinfo{volume}{39} (\bibinfo{year}{2023}) \bibinfo{pages}{btad671}.
  \DOIprefix\doi{10.1093/bioinformatics/btad671}.
%Type = Article
\bibitem[{Yang et~al.(2023)Yang, Zanichelli, and Yeh}]{Yang2023}
\bibinfo{author}{K.~K. Yang}, \bibinfo{author}{N.~Zanichelli},
  \bibinfo{author}{H.~Yeh},
\newblock \bibinfo{title}{Masked inverse folding with sequence transfer for
  protein representation learning},
\newblock \bibinfo{journal}{Protein engineering, design \& selection: PEDS}
  \bibinfo{volume}{36} (\bibinfo{year}{2023}) \bibinfo{pages}{gzad015}.
  \DOIprefix\doi{10.1093/protein/gzad015}.
%Type = Article
\bibitem[{Jiang et~al.(2024)Jiang, Li, Dong, Yu, Sun, Wu, Huang, Kang, Pei,
  Zhang, Wang, Xu, Xin, Ouyang, Fan, Zheng, Tan, Hu, Xiong, Feng, Yang, Liu,
  Song, Liu, Hong, and Tan}]{Jiang2024}
\bibinfo{author}{F.~Jiang}, \bibinfo{author}{M.~Li}, \bibinfo{author}{J.~Dong},
  \bibinfo{author}{Y.~Yu}, \bibinfo{author}{X.~Sun}, \bibinfo{author}{B.~Wu},
  \bibinfo{author}{J.~Huang}, \bibinfo{author}{L.~Kang},
  \bibinfo{author}{Y.~Pei}, \bibinfo{author}{L.~Zhang},
  \bibinfo{author}{S.~Wang}, \bibinfo{author}{W.~Xu}, \bibinfo{author}{J.~Xin},
  \bibinfo{author}{W.~Ouyang}, \bibinfo{author}{G.~Fan},
  \bibinfo{author}{L.~Zheng}, \bibinfo{author}{Y.~Tan},
  \bibinfo{author}{Z.~Hu}, \bibinfo{author}{Y.~Xiong},
  \bibinfo{author}{Y.~Feng}, \bibinfo{author}{G.~Yang},
  \bibinfo{author}{Q.~Liu}, \bibinfo{author}{J.~Song},
  \bibinfo{author}{J.~Liu}, \bibinfo{author}{L.~Hong},
  \bibinfo{author}{P.~Tan},
\newblock \bibinfo{title}{A general temperature-guided language model to design
  proteins of enhanced stability and activity},
\newblock \bibinfo{journal}{Science Advances} \bibinfo{volume}{10}
  (\bibinfo{year}{2024}) \bibinfo{pages}{eadr2641}. \URLprefix
  \url{https://www.science.org/doi/10.1126/sciadv.adr2641}.
  \DOIprefix\doi{10.1126/sciadv.adr2641}.
%Type = Article
\bibitem[{Li et~al.(2024)Li, Yao, and Fan}]{Li2024a}
\bibinfo{author}{G.~Li}, \bibinfo{author}{S.~Yao}, \bibinfo{author}{L.~Fan},
\newblock \bibinfo{title}{{ProSTAGE}: {Predicting} {Effects} of {Mutations} on
  {Protein} {Stability} by {Using} {Protein} {Embeddings} and {Graph}
  {Convolutional} {Networks}},
\newblock \bibinfo{journal}{Journal of Chemical Information and Modeling}
  \bibinfo{volume}{64} (\bibinfo{year}{2024}) \bibinfo{pages}{340--347}.
  \URLprefix \url{https://doi.org/10.1021/acs.jcim.3c01697}.
  \DOIprefix\doi{10.1021/acs.jcim.3c01697}.
%Type = Article
\bibitem[{Cuturello et~al.(2024)Cuturello, Celoria, Ansuini, and
  Cazzaniga}]{Cuturello2024}
\bibinfo{author}{F.~Cuturello}, \bibinfo{author}{M.~Celoria},
  \bibinfo{author}{A.~Ansuini}, \bibinfo{author}{A.~Cazzaniga},
\newblock \bibinfo{title}{Enhancing predictions of protein stability changes
  induced by single mutations using {MSA}-based language models},
\newblock \bibinfo{journal}{Bioinformatics} \bibinfo{volume}{40}
  (\bibinfo{year}{2024}) \bibinfo{pages}{btae447}. \URLprefix
  \url{https://doi.org/10.1093/bioinformatics/btae447}.
  \DOIprefix\doi{10.1093/bioinformatics/btae447}.
%Type = Article
\bibitem[{Dieckhaus et~al.(2024)Dieckhaus, Brocidiacono, Randolph, and
  Kuhlman}]{Dieckhaus2024}
\bibinfo{author}{H.~Dieckhaus}, \bibinfo{author}{M.~Brocidiacono},
  \bibinfo{author}{N.~Z. Randolph}, \bibinfo{author}{B.~Kuhlman},
\newblock \bibinfo{title}{Transfer learning to leverage larger datasets for
  improved prediction of protein stability changes},
\newblock \bibinfo{journal}{Proceedings of the National Academy of Sciences}
  \bibinfo{volume}{121} (\bibinfo{year}{2024}) \bibinfo{pages}{e2314853121}.
  \URLprefix \url{https://www.pnas.org/doi/10.1073/pnas.2314853121}.
  \DOIprefix\doi{10.1073/pnas.2314853121}.
%Type = Article
\bibitem[{Chu et~al.(2024)Chu, Narang, and Siegel}]{Chu2024}
\bibinfo{author}{S.~K.~S. Chu}, \bibinfo{author}{K.~Narang},
  \bibinfo{author}{J.~B. Siegel},
\newblock \bibinfo{title}{Protein stability prediction by fine-tuning a protein
  language model on a mega-scale dataset},
\newblock \bibinfo{journal}{PLoS computational biology} \bibinfo{volume}{20}
  (\bibinfo{year}{2024}) \bibinfo{pages}{e1012248}.
  \DOIprefix\doi{10.1371/journal.pcbi.1012248}.
%Type = Article
\bibitem[{Sun et~al.(2025)Sun, Zhu, Cui, and Wu}]{Sun2025}
\bibinfo{author}{J.~Sun}, \bibinfo{author}{T.~Zhu}, \bibinfo{author}{Y.~Cui},
  \bibinfo{author}{B.~Wu},
\newblock \bibinfo{title}{Structure-based self-supervised learning enables
  ultrafast protein stability prediction upon mutation},
\newblock \bibinfo{journal}{Innovation (Cambridge (Mass.))} \bibinfo{volume}{6}
  (\bibinfo{year}{2025}) \bibinfo{pages}{100750}.
  \DOIprefix\doi{10.1016/j.xinn.2024.100750}.
%Type = Misc
\bibitem[{Chen et~al.(2025)Chen, Wang, Hu, Li, Qian, and Song}]{plm2025}
\bibinfo{author}{J.-Y. Chen}, \bibinfo{author}{J.-F. Wang},
  \bibinfo{author}{Y.~Hu}, \bibinfo{author}{X.-H. Li}, \bibinfo{author}{Y.-R.
  Qian}, \bibinfo{author}{C.-L. Song}, \bibinfo{title}{A {Comprehensive}
  {Review} of {Protein} {Language} {Models}}, \bibinfo{year}{2025}. \URLprefix
  \url{http://arxiv.org/abs/2502.06881}.
  \DOIprefix\doi{10.48550/arXiv.2502.06881}, \bibinfo{note}{arXiv:2502.06881
  [q-bio] version: 1}.
%Type = Article
\bibitem[{Ruffolo and Madani(2024)}]{Ruffolo2024}
\bibinfo{author}{J.~A. Ruffolo}, \bibinfo{author}{A.~Madani},
\newblock \bibinfo{title}{Designing proteins with language models},
\newblock \bibinfo{journal}{Nature Biotechnology} \bibinfo{volume}{42}
  (\bibinfo{year}{2024}) \bibinfo{pages}{200--202}. \URLprefix
  \url{https://www.nature.com/articles/s41587-024-02123-4}.
  \DOIprefix\doi{10.1038/s41587-024-02123-4}.
%Type = Article
\bibitem[{Savojardo et~al.(2025)Savojardo, Manfredi, Martelli, and
  Casadio}]{savojardo2025ddgemb}
\bibinfo{author}{C.~Savojardo}, \bibinfo{author}{M.~Manfredi},
  \bibinfo{author}{P.~L. Martelli}, \bibinfo{author}{R.~Casadio},
\newblock \bibinfo{title}{Ddgemb: predicting protein stability change upon
  single-and multi-point variations with embeddings and deep learning},
\newblock \bibinfo{journal}{Bioinformatics}  (\bibinfo{year}{2025})
  \bibinfo{pages}{btaf019}.
%Type = Article
\bibitem[{Pucci et~al.(2018)Pucci, Bernaerts, Kwasigroch, and
  Rooman}]{pucci2018quantification}
\bibinfo{author}{F.~Pucci}, \bibinfo{author}{K.~V. Bernaerts},
  \bibinfo{author}{J.~M. Kwasigroch}, \bibinfo{author}{M.~Rooman},
\newblock \bibinfo{title}{Quantification of biases in predictions of protein
  stability changes upon mutations},
\newblock \bibinfo{journal}{Bioinformatics} \bibinfo{volume}{34}
  (\bibinfo{year}{2018}) \bibinfo{pages}{3659--3665}.
%Type = Article
\bibitem[{Pancotti et~al.(2022)Pancotti, Benevenuta, Birolo, Alberini, Repetto,
  Sanavia, Capriotti, and Fariselli}]{pancotti2022b}
\bibinfo{author}{C.~Pancotti}, \bibinfo{author}{S.~Benevenuta},
  \bibinfo{author}{G.~Birolo}, \bibinfo{author}{V.~Alberini},
  \bibinfo{author}{V.~Repetto}, \bibinfo{author}{T.~Sanavia},
  \bibinfo{author}{E.~Capriotti}, \bibinfo{author}{P.~Fariselli},
\newblock \bibinfo{title}{Predicting protein stability changes upon
  single-point mutation: a thorough comparison of the available tools on a new
  dataset},
\newblock \bibinfo{journal}{Briefings in Bioinformatics} \bibinfo{volume}{23}
  (\bibinfo{year}{2022}) \bibinfo{pages}{bbab555}.
%Type = Article
\bibitem[{Reeves and Kalyaanamoorthy(2024)}]{reeves2024zero}
\bibinfo{author}{S.~Reeves}, \bibinfo{author}{S.~Kalyaanamoorthy},
\newblock \bibinfo{title}{Zero-shot transfer of protein sequence likelihood
  models to thermostability prediction},
\newblock \bibinfo{journal}{Nature Machine Intelligence} \bibinfo{volume}{6}
  (\bibinfo{year}{2024}) \bibinfo{pages}{1063--1076}.
%Type = Article
\bibitem[{Rollo et~al.(2023)Rollo, Pancotti, Birolo, Rossi, Sanavia, and
  Fariselli}]{rollo2023influence}
\bibinfo{author}{C.~Rollo}, \bibinfo{author}{C.~Pancotti},
  \bibinfo{author}{G.~Birolo}, \bibinfo{author}{I.~Rossi},
  \bibinfo{author}{T.~Sanavia}, \bibinfo{author}{P.~Fariselli},
\newblock \bibinfo{title}{Influence of model structures on predictors of
  protein stability changes from single-point mutations},
\newblock \bibinfo{journal}{Genes} \bibinfo{volume}{14} (\bibinfo{year}{2023})
  \bibinfo{pages}{2228}.
%Type = Article
\bibitem[{Jumper et~al.(2021)Jumper, Evans, Pritzel, Green, Figurnov,
  Ronneberger, Tunyasuvunakool, Bates, Žídek, Potapenko, Bridgland, Meyer,
  Kohl, Ballard, Cowie, Romera-Paredes, Nikolov, Jain, Adler, Back, Petersen,
  Reiman, Clancy, Zielinski, Steinegger, Pacholska, Berghammer, Bodenstein,
  Silver, Vinyals, Senior, Kavukcuoglu, Kohli, and Hassabis}]{Jumper2021}
\bibinfo{author}{J.~Jumper}, \bibinfo{author}{R.~Evans},
  \bibinfo{author}{A.~Pritzel}, \bibinfo{author}{T.~Green},
  \bibinfo{author}{M.~Figurnov}, \bibinfo{author}{O.~Ronneberger},
  \bibinfo{author}{K.~Tunyasuvunakool}, \bibinfo{author}{R.~Bates},
  \bibinfo{author}{A.~Žídek}, \bibinfo{author}{A.~Potapenko},
  \bibinfo{author}{A.~Bridgland}, \bibinfo{author}{C.~Meyer},
  \bibinfo{author}{S.~A.~A. Kohl}, \bibinfo{author}{A.~J. Ballard},
  \bibinfo{author}{A.~Cowie}, \bibinfo{author}{B.~Romera-Paredes},
  \bibinfo{author}{S.~Nikolov}, \bibinfo{author}{R.~Jain},
  \bibinfo{author}{J.~Adler}, \bibinfo{author}{T.~Back},
  \bibinfo{author}{S.~Petersen}, \bibinfo{author}{D.~Reiman},
  \bibinfo{author}{E.~Clancy}, \bibinfo{author}{M.~Zielinski},
  \bibinfo{author}{M.~Steinegger}, \bibinfo{author}{M.~Pacholska},
  \bibinfo{author}{T.~Berghammer}, \bibinfo{author}{S.~Bodenstein},
  \bibinfo{author}{D.~Silver}, \bibinfo{author}{O.~Vinyals},
  \bibinfo{author}{A.~W. Senior}, \bibinfo{author}{K.~Kavukcuoglu},
  \bibinfo{author}{P.~Kohli}, \bibinfo{author}{D.~Hassabis},
\newblock \bibinfo{title}{Highly accurate protein structure prediction with
  {AlphaFold}},
\newblock \bibinfo{journal}{Nature} \bibinfo{volume}{596}
  (\bibinfo{year}{2021}) \bibinfo{pages}{583--589}. \URLprefix
  \url{https://www.nature.com/articles/s41586-021-03819-2}.
  \DOIprefix\doi{10.1038/s41586-021-03819-2}.
%Type = Article
\bibitem[{Baek et~al.(2021)Baek, DiMaio, Anishchenko, Dauparas, Ovchinnikov,
  Lee, Wang, Cong, Kinch, Schaeffer, Millán, Park, Adams, Glassman,
  DeGiovanni, Pereira, Rodrigues, van Dijk, Ebrecht, Opperman, Sagmeister,
  Buhlheller, Pavkov-Keller, Rathinaswamy, Dalwadi, Yip, Burke, Garcia,
  Grishin, Adams, Read, and Baker}]{Baek2021}
\bibinfo{author}{M.~Baek}, \bibinfo{author}{F.~DiMaio},
  \bibinfo{author}{I.~Anishchenko}, \bibinfo{author}{J.~Dauparas},
  \bibinfo{author}{S.~Ovchinnikov}, \bibinfo{author}{G.~R. Lee},
  \bibinfo{author}{J.~Wang}, \bibinfo{author}{Q.~Cong}, \bibinfo{author}{L.~N.
  Kinch}, \bibinfo{author}{R.~D. Schaeffer}, \bibinfo{author}{C.~Millán},
  \bibinfo{author}{H.~Park}, \bibinfo{author}{C.~Adams}, \bibinfo{author}{C.~R.
  Glassman}, \bibinfo{author}{A.~DeGiovanni}, \bibinfo{author}{J.~H. Pereira},
  \bibinfo{author}{A.~V. Rodrigues}, \bibinfo{author}{A.~A. van Dijk},
  \bibinfo{author}{A.~C. Ebrecht}, \bibinfo{author}{D.~J. Opperman},
  \bibinfo{author}{T.~Sagmeister}, \bibinfo{author}{C.~Buhlheller},
  \bibinfo{author}{T.~Pavkov-Keller}, \bibinfo{author}{M.~K. Rathinaswamy},
  \bibinfo{author}{U.~Dalwadi}, \bibinfo{author}{C.~K. Yip},
  \bibinfo{author}{J.~E. Burke}, \bibinfo{author}{K.~C. Garcia},
  \bibinfo{author}{N.~V. Grishin}, \bibinfo{author}{P.~D. Adams},
  \bibinfo{author}{R.~J. Read}, \bibinfo{author}{D.~Baker},
\newblock \bibinfo{title}{Accurate prediction of protein structures and
  interactions using a three-track neural network},
\newblock \bibinfo{journal}{Science} \bibinfo{volume}{373}
  (\bibinfo{year}{2021}) \bibinfo{pages}{871--876}. \URLprefix
  \url{https://www.science.org/doi/10.1126/science.abj8754}.
  \DOIprefix\doi{10.1126/science.abj8754}.
%Type = Article
\bibitem[{Ahdritz et~al.(2024)Ahdritz, Bouatta, Floristean, Kadyan, Xia,
  Gerecke, O’Donnell, Berenberg, Fisk, Zanichelli
  et~al.}]{ahdritz2024openfold}
\bibinfo{author}{G.~Ahdritz}, \bibinfo{author}{N.~Bouatta},
  \bibinfo{author}{C.~Floristean}, \bibinfo{author}{S.~Kadyan},
  \bibinfo{author}{Q.~Xia}, \bibinfo{author}{W.~Gerecke},
  \bibinfo{author}{T.~J. O’Donnell}, \bibinfo{author}{D.~Berenberg},
  \bibinfo{author}{I.~Fisk}, \bibinfo{author}{N.~Zanichelli}, et~al.,
\newblock \bibinfo{title}{Openfold: Retraining alphafold2 yields new insights
  into its learning mechanisms and capacity for generalization},
\newblock \bibinfo{journal}{Nature Methods} \bibinfo{volume}{21}
  (\bibinfo{year}{2024}) \bibinfo{pages}{1514--1524}.
%Type = Misc
\bibitem[{Ouyang-Zhang et~al.(2023)Ouyang-Zhang, Diaz, Klivans, and
  Krähenbühl}]{mutateeverything2023}
\bibinfo{author}{J.~Ouyang-Zhang}, \bibinfo{author}{D.~J. Diaz},
  \bibinfo{author}{A.~R. Klivans}, \bibinfo{author}{P.~Krähenbühl},
  \bibinfo{title}{Predicting a {Protein}'s {Stability} under a {Million}
  {Mutations}}, \bibinfo{year}{2023}. \URLprefix
  \url{http://arxiv.org/abs/2310.12979}.
  \DOIprefix\doi{10.48550/arXiv.2310.12979}, \bibinfo{note}{arXiv:2310.12979
  [q-bio]}.
%Type = Article
\bibitem[{Dieckhaus and Kuhlman(2025)}]{Dieckhaus2025}
\bibinfo{author}{H.~Dieckhaus}, \bibinfo{author}{B.~Kuhlman},
\newblock \bibinfo{title}{Protein stability models fail to capture epistatic
  interactions of double point mutations},
\newblock \bibinfo{journal}{Protein Science} \bibinfo{volume}{34}
  (\bibinfo{year}{2025}) \bibinfo{pages}{e70003}. \URLprefix
  \url{https://onlinelibrary.wiley.com/doi/abs/10.1002/pro.70003}.
  \DOIprefix\doi{10.1002/pro.70003}.
%Type = Article
\bibitem[{Montanucci et~al.(2022)Montanucci, Capriotti, Birolo, Benevenuta,
  Pancotti, Lal, and Fariselli}]{montanucci2022ddgun}
\bibinfo{author}{L.~Montanucci}, \bibinfo{author}{E.~Capriotti},
  \bibinfo{author}{G.~Birolo}, \bibinfo{author}{S.~Benevenuta},
  \bibinfo{author}{C.~Pancotti}, \bibinfo{author}{D.~Lal},
  \bibinfo{author}{P.~Fariselli},
\newblock \bibinfo{title}{Ddgun: an untrained predictor of protein stability
  changes upon amino acid variants},
\newblock \bibinfo{journal}{Nucleic Acids Research} \bibinfo{volume}{50}
  (\bibinfo{year}{2022}) \bibinfo{pages}{W222--W227}.
%Type = Article
\bibitem[{Chen et~al.(2024)Chen, Xu, Liu, Xing, and Gong}]{chen2024}
\bibinfo{author}{Y.~Chen}, \bibinfo{author}{Y.~Xu}, \bibinfo{author}{D.~Liu},
  \bibinfo{author}{Y.~Xing}, \bibinfo{author}{H.~Gong},
\newblock \bibinfo{title}{An end-to-end framework for the prediction of protein
  structure and fitness from single sequence},
\newblock \bibinfo{journal}{Nature Communications} \bibinfo{volume}{15}
  (\bibinfo{year}{2024}) \bibinfo{pages}{7400}. \URLprefix
  \url{https://www.nature.com/articles/s41467-024-51776-x}.
  \DOIprefix\doi{10.1038/s41467-024-51776-x}.
%Type = Article
\bibitem[{Lin et~al.(2022)Lin, Akin, Rao, Hie, Zhu, Lu, Smetanin, dos
  Santos~Costa, Fazel-Zarandi, Sercu, Candido et~al.}]{lin2022language}
\bibinfo{author}{Z.~Lin}, \bibinfo{author}{H.~Akin}, \bibinfo{author}{R.~Rao},
  \bibinfo{author}{B.~Hie}, \bibinfo{author}{Z.~Zhu}, \bibinfo{author}{W.~Lu},
  \bibinfo{author}{N.~Smetanin}, \bibinfo{author}{A.~dos Santos~Costa},
  \bibinfo{author}{M.~Fazel-Zarandi}, \bibinfo{author}{T.~Sercu},
  \bibinfo{author}{S.~Candido}, et~al.,
\newblock \bibinfo{title}{Language models of protein sequences at the scale of
  evolution enable accurate structure prediction},
\newblock \bibinfo{journal}{bioRxiv}  (\bibinfo{year}{2022}).
%Type = Article
\bibitem[{Samaga et~al.(2021)Samaga, Raghunathan, and Priyakumar}]{Samaga2021}
\bibinfo{author}{Y.~B.~L. Samaga}, \bibinfo{author}{S.~Raghunathan},
  \bibinfo{author}{U.~D. Priyakumar},
\newblock \bibinfo{title}{{SCONES}: {Self}-{Consistent} {Neural} {Network} for
  {Protein} {Stability} {Prediction} {Upon} {Mutation}},
\newblock \bibinfo{journal}{The Journal of Physical Chemistry B}
  \bibinfo{volume}{125} (\bibinfo{year}{2021}) \bibinfo{pages}{10657--10671}.
  \URLprefix \url{https://doi.org/10.1021/acs.jpcb.1c04913}.
  \DOIprefix\doi{10.1021/acs.jpcb.1c04913}.
%Type = Unpublished
\bibitem[{{Schr\"odinger, LLC}(2015)}]{PyMOL}
\bibinfo{author}{{Schr\"odinger, LLC}}, \bibinfo{title}{The {PyMOL} molecular
  graphics system, version~1.8}, \bibinfo{year}{2015}.
%Type = Article
\bibitem[{Capriotti et~al.(2008)Capriotti, Fariselli, Rossi, and
  Casadio}]{capriotti2008three}
\bibinfo{author}{E.~Capriotti}, \bibinfo{author}{P.~Fariselli},
  \bibinfo{author}{I.~Rossi}, \bibinfo{author}{R.~Casadio},
\newblock \bibinfo{title}{A three-state prediction of single point mutations on
  protein stability changes},
\newblock \bibinfo{journal}{BMC bioinformatics} \bibinfo{volume}{9}
  (\bibinfo{year}{2008}) \bibinfo{pages}{1--9}.
%Type = Article
\bibitem[{Diaz et~al.(2024)Diaz, Gong, Ouyang-Zhang, Loy, Wells, Yang,
  Ellington, Dimakis, and Klivans}]{Diaz2024}
\bibinfo{author}{D.~J. Diaz}, \bibinfo{author}{C.~Gong},
  \bibinfo{author}{J.~Ouyang-Zhang}, \bibinfo{author}{J.~M. Loy},
  \bibinfo{author}{J.~Wells}, \bibinfo{author}{D.~Yang}, \bibinfo{author}{A.~D.
  Ellington}, \bibinfo{author}{A.~G. Dimakis}, \bibinfo{author}{A.~R. Klivans},
\newblock \bibinfo{title}{Stability {Oracle}: a structure-based
  graph-transformer framework for identifying stabilizing mutations},
\newblock \bibinfo{journal}{Nature Communications} \bibinfo{volume}{15}
  (\bibinfo{year}{2024}) \bibinfo{pages}{6170}. \URLprefix
  \url{https://www.nature.com/articles/s41467-024-49780-2}.
  \DOIprefix\doi{10.1038/s41467-024-49780-2}.
%Type = Article
\bibitem[{Nair and Vihinen(2013)}]{nair2013v}
\bibinfo{author}{P.~S. Nair}, \bibinfo{author}{M.~Vihinen},
\newblock \bibinfo{title}{V ari b ench: A benchmark database for variations},
\newblock \bibinfo{journal}{Human mutation} \bibinfo{volume}{34}
  (\bibinfo{year}{2013}) \bibinfo{pages}{42--49}.
%Type = Article
\bibitem[{Xavier et~al.(2021)Xavier, Nguyen, Karmarkar, Portelli, Rezende,
  Velloso, Ascher, and Pires}]{xavier2021thermomutdb}
\bibinfo{author}{J.~S. Xavier}, \bibinfo{author}{T.-B. Nguyen},
  \bibinfo{author}{M.~Karmarkar}, \bibinfo{author}{S.~Portelli},
  \bibinfo{author}{P.~M. Rezende}, \bibinfo{author}{J.~P. Velloso},
  \bibinfo{author}{D.~B. Ascher}, \bibinfo{author}{D.~E. Pires},
\newblock \bibinfo{title}{Thermomutdb: a thermodynamic database for missense
  mutations},
\newblock \bibinfo{journal}{Nucleic acids research} \bibinfo{volume}{49}
  (\bibinfo{year}{2021}) \bibinfo{pages}{D475--D479}.
%Type = Article
\bibitem[{Nikam et~al.(2021)Nikam, Kulandaisamy, Harini, Sharma, and
  Gromiha}]{nikam2021prothermdb}
\bibinfo{author}{R.~Nikam}, \bibinfo{author}{A.~Kulandaisamy},
  \bibinfo{author}{K.~Harini}, \bibinfo{author}{D.~Sharma},
  \bibinfo{author}{M.~M. Gromiha},
\newblock \bibinfo{title}{Prothermdb: thermodynamic database for proteins and
  mutants revisited after 15 years},
\newblock \bibinfo{journal}{Nucleic acids research} \bibinfo{volume}{49}
  (\bibinfo{year}{2021}) \bibinfo{pages}{D420--D424}.
%Type = Article
\bibitem[{Montanucci et~al.(2019)Montanucci, Capriotti, Frank, Ben-Tal, and
  Fariselli}]{montanucci2019ddgun}
\bibinfo{author}{L.~Montanucci}, \bibinfo{author}{E.~Capriotti},
  \bibinfo{author}{Y.~Frank}, \bibinfo{author}{N.~Ben-Tal},
  \bibinfo{author}{P.~Fariselli},
\newblock \bibinfo{title}{Ddgun: an untrained method for the prediction of
  protein stability changes upon single and multiple point variations},
\newblock \bibinfo{journal}{BMC bioinformatics} \bibinfo{volume}{20}
  (\bibinfo{year}{2019}) \bibinfo{pages}{1--10}.
%Type = Article
\bibitem[{Bava et~al.(2004)Bava, Gromiha, Uedaira, Kitajima, and
  Sarai}]{bava2004protherm}
\bibinfo{author}{K.~A. Bava}, \bibinfo{author}{M.~M. Gromiha},
  \bibinfo{author}{H.~Uedaira}, \bibinfo{author}{K.~Kitajima},
  \bibinfo{author}{A.~Sarai},
\newblock \bibinfo{title}{Protherm, version 4.0: thermodynamic database for
  proteins and mutants},
\newblock \bibinfo{journal}{Nucleic acids research} \bibinfo{volume}{32}
  (\bibinfo{year}{2004}) \bibinfo{pages}{D120--D121}.
%Type = Inproceedings
\bibitem[{Waswani et~al.(2017)Waswani, Shazeer, Parmar, Uszkoreit, Jones,
  Gomez, Kaiser, and Polosukhin}]{waswani2017attention}
\bibinfo{author}{A.~Waswani}, \bibinfo{author}{N.~Shazeer},
  \bibinfo{author}{N.~Parmar}, \bibinfo{author}{J.~Uszkoreit},
  \bibinfo{author}{L.~Jones}, \bibinfo{author}{A.~Gomez},
  \bibinfo{author}{L.~Kaiser}, \bibinfo{author}{I.~Polosukhin},
\newblock \bibinfo{title}{Attention is all you need},
\newblock in: \bibinfo{booktitle}{NIPS}, \bibinfo{year}{2017}.
%Type = Article
\bibitem[{Steinegger and S{\"o}ding(2017)}]{steinegger2017mmseqs2}
\bibinfo{author}{M.~Steinegger}, \bibinfo{author}{J.~S{\"o}ding},
\newblock \bibinfo{title}{Mmseqs2 enables sensitive protein sequence searching
  for the analysis of massive data sets},
\newblock \bibinfo{journal}{Nature biotechnology} \bibinfo{volume}{35}
  (\bibinfo{year}{2017}) \bibinfo{pages}{1026--1028}.

\end{thebibliography}

%% Forza di nuovo la stampa di eventuali figure/tabelle residue
\clearpage
\onecolumn
%% Include the supplementary section dopo tutto
\section{Supplementary Information}
\label{sec:results}

\subsection{Comparison Between Janus Base, JanusAsym, and JanusFineTuned}
\label{sec:comparison_3janus}

In this chapter, we present a comparative analysis of the JanusDDG Base, JanusDDG only Antisym., and JanusDDG models, evaluating their performance on blind test sets. The assessment includes both single and multiple mutations, providing a comprehensive overview of how each model handles different mutation scenarios. JanusDDG only Antisym. is the antisymmetric model by construction, derived from the JanusDDG Base to enforce antisymmetry in its predictions. JanusDDG is the fine-tuned version of the base model.

\begin{table*}[h]
    \centering
    \caption{Comparison Between Janus Base, JanusAsym, and JanusFineTuned on the S669 independent test set of single-point variations.}
    \label{tab:metrics_single}
    \renewcommand{\arraystretch}{1.2}
    \resizebox{\textwidth}{!}{%
    \begin{tabular}{l c ccc ccc ccc c c}
        \hline
        \textbf{Method} & \textbf{Input} & \multicolumn{3}{c}{\textbf{Total}} & \multicolumn{3}{c}{\textbf{Direct}} & \multicolumn{3}{c}{\textbf{Reverse}} & \textbf{rd-r} & $\langle \delta \rangle$ \\
        & & PCC & RMSE & MAE & PCC & RMSE & MAE & PCC & RMSE & MAE & & \\
        \hline
        JanusDDG & SEQ & \textbf{0.69} & 1.39 & 0.97 & \textbf{0.55} & 1.39 & 0.97 & \textbf{0.55} & 1.39 & 0.97 & \textbf{-1} & \textbf{0.00} \\
        JanusDDG only Antisym. & SEQ & \textbf{0.69} & \textbf{1.37} & \textbf{0.96} & \textbf{0.55} & \textbf{1.37} & \textbf{0.96} & \textbf{0.55} & \textbf{1.37} & \textbf{0.96} & \textbf{-1} & \textbf{0.00} \\
        JanusDDG Base & SEQ & \textbf{0.69} & 1.38 & \textbf{0.96} & \textbf{0.55} & \textbf{1.37} & \textbf{0.96} & 0.53 & 1.39 & 0.97 & -0.95 & 0.02 \\
             
        \hline
    \end{tabular}%
    }
\end{table*}

\begin{table*}[h]
    \centering
    \caption{Comparison Between Janus Base, JanusAsym, and JanusFineTuned on the S461 independent test set of single-point variations.}
    \label{tab:metrics_single}
    \renewcommand{\arraystretch}{1.2}
    \resizebox{\textwidth}{!}{%
    \begin{tabular}{l c ccc ccc ccc c c}
        \hline
        \textbf{Method} & \textbf{Input} & \multicolumn{3}{c}{\textbf{Total}} & \multicolumn{3}{c}{\textbf{Direct}} & \multicolumn{3}{c}{\textbf{Reverse}} & \textbf{rd-r} & $\langle \delta \rangle$ \\
        & & PCC & RMSE & MAE & PCC & RMSE & MAE & PCC & RMSE & MAE & & \\
        \hline
        JanusDDG FineTuned & SEQ & 0.83 & 0.96 & 0.71 & 0.69 & 0.96 & 0.71 & 0.69 & 0.96 & 0.71 & \textbf{-1} & \textbf{0.00} \\
        JanusDDG only Antisym. & SEQ & \textbf{0.84} & \textbf{0.92} & \textbf{0.70} & \textbf{0.70} & \textbf{0.92} & \textbf{0.70} & \textbf{0.70} & \textbf{0.92} & \textbf{0.70} & \textbf{-1} & \textbf{0.00} \\
        JanusDDG Base & SEQ & \textbf{0.84} & 0.93 & 0.71 & \textbf{0.70} & \textbf{0.92} & \textbf{0.70} & 0.68 & 0.95 & 0.72 & -0.95 & \textbf{0.00} \\
             
        \hline
    \end{tabular}%
    }
\end{table*}

\begin{table*}[h]
    \centering
    \caption{Comparison Between Janus Base, JanusAsym, and JanusFineTuned on the PTmul-NR independent test set of multi-point variations.}
    \label{tab:metrics_single}
    \renewcommand{\arraystretch}{1.2}
    \resizebox{\textwidth}{!}{%
    \begin{tabular}{l c ccc ccc ccc c c}
        \hline
        \textbf{Method} & \textbf{Input} & \multicolumn{3}{c}{\textbf{Total}} & \multicolumn{3}{c}{\textbf{Direct}} & \multicolumn{3}{c}{\textbf{Reverse}} & \textbf{rd-r} & $\langle \delta \rangle$ \\
        & & PCC & RMSE & MAE & PCC & RMSE & MAE & PCC & RMSE & MAE & & \\
        \hline
        JanusDDG FineTuned & SEQ & \textbf{0.61} & \textbf{2.06} & \textbf{1.57} & \textbf{0.61} & \textbf{2.06} & \textbf{1.57} & \textbf{0.61} & \textbf{2.06} & \textbf{1.57} & \textbf{-1} & \textbf{0.00} \\
        JanusDDG only Antisym. & SEQ & 0.55 & 2.17 & 1.67 & 0.55 & 2.17 & 1.67 & 0.55 & 2.17 & 1.67 & \textbf{-1} & \textbf{0.00} \\
        JanusDDG Base & SEQ & 0.54 & 2.18 & 1.67 & 0.55 & 2.17 & 1.63 & 0.54 & 2.19 & 1.71 & -0.96 & -0.09 \\
             
        \hline
    \end{tabular}%
    }
\end{table*}

\newpage

\subsection{Performance of JanusDDG Across Benchmark Datasets}

%In this section, we present the tables with the performance of JanusDDG and other models from the literature on blind test sets.

\begin{table*}[h]
    \centering
    \caption{Comparative benchmark of different sequence- and structure-based methods on the S669 independent test set of single-point variations. The models’ performance data, excluding JanusDDG, were taken from~\cite{savojardo2025ddgemb}.}
    \label{tab:metrics_single}
    \renewcommand{\arraystretch}{1.2}
    \resizebox{\textwidth}{!}{%
    \begin{tabular}{l c ccc ccc ccc c c}
        \hline
        \textbf{Method} & \textbf{Input} & \multicolumn{3}{c}{\textbf{Total}} & \multicolumn{3}{c}{\textbf{Direct}} & \multicolumn{3}{c}{\textbf{Reverse}} & \textbf{Pearson d-r} & $\langle \delta \rangle$ \\
        & & PCC & RMSE & MAE & PCC & RMSE & MAE & PCC & RMSE & MAE & & \\
        \hline
        \textbf{JanusDDG} & SEQ & \textbf{0.69} & \textbf{1.39} & \textbf{0.97} & \textbf{0.55} & \textbf{1.39} & \textbf{0.97} & \textbf{0.55} & \textbf{1.39} & \textbf{0.97} & \textbf{-1} & \textbf{0.00} \\
        DDGemb & SEQ & 0.68 & 1.40 & 0.99 & 0.53 & 1.40 & 0.99 & 0.52 & 1.40 & 0.99 & -0.97 & 0.01 \\
        PROSTATA & SEQ & 0.65 & 1.45 & 1.00 & 0.49 & 1.45 & 1.00 & 0.49 & 1.45 & 0.99 & -0.99 & -0.01 \\
        ACDC-NN & 3D & 0.61 & 1.50 & 1.05 & 0.46 & 1.49 & 1.05 & 0.45 & 1.50 & 1.06 & -0.98 & 0.02 \\
        INPS-Seq & SEQ & 0.61 & 1.52 & 1.10 & 0.43 & 1.52 & 1.09 & 0.43 & 1.53 & 1.10 & -1.00 & 0.00 \\
        PremPS & 3D & 0.62 & 1.49 & 1.07 & 0.41 & 1.50 & 1.08 & 0.42 & 1.49 & 1.05 & -0.85 & 0.09 \\
        ACDC-NN-Seq & SEQ & 0.59 & 1.53 & 1.08 & 0.42 & 1.53 & 1.08 & 0.42 & 1.53 & 1.08 & -1.00 & 0.00 \\
        DDGun3D & 3D & 0.57 & 1.61 & 1.13 & 0.43 & 1.60 & 1.11 & 0.41 & 1.62 & 1.14 & -0.97 & 0.05 \\
        INPS3D & 3D & 0.55 & 1.64 & 1.19 & 0.43 & 1.50 & 1.07 & 0.33 & 1.77 & 1.31 & -0.50 & 0.38 \\
        THPLM & SEQ & 0.53 & 1.63 & - & 0.39 & 1.60 & - & 0.35 & 1.66 & - & -0.96 & -0.01 \\
        ThermoNet & 3D & 0.51 & 1.64 & 1.20 & 0.39 & 1.62 & 1.17 & 0.38 & 1.66 & 1.23 & -0.85 & 0.05 \\
        DDGun & SEQ & 0.57 & 1.74 & 1.25 & 0.41 & 1.72 & 1.25 & 0.38 & 1.75 & 1.25 & -0.96 & 0.05 \\
        MAESTRO & 3D & 0.44 & 1.80 & 1.30 & 0.50 & 1.44 & 1.06 & 0.20 & 2.10 & 1.66 & -0.22 & 0.57 \\
        ThermoMPNN & SEQ & 0.43 & 1.52 & - & - & - & - & - & - & - & - & - \\
        Dynamut & 3D & 0.50 & 1.65 & 1.21 & 0.41 & 1.60 & 1.19 & 0.34 & 1.69 & 1.24 & -0.58 & 0.06 \\
        PoPMuSiC & 3D & 0.46 & 1.82 & 1.37 & 0.41 & 1.51 & 1.09 & 0.24 & 2.09 & 1.64 & -0.32 & 0.69 \\
        DUET & 3D & 0.41 & 1.86 & 1.39 & 0.41 & 1.52 & 1.10 & 0.23 & 2.14 & 1.68 & -0.12 & 0.67 \\
        I-Mutant3.0-Seq & SEQ & 0.37 & 1.91 & 1.47 & 0.34 & 1.54 & 1.15 & 0.22 & 2.22 & 1.79 & -0.48 & 0.76 \\
        SDM & 3D & 0.32 & 1.93 & 1.45 & 0.41 & 1.67 & 1.26 & 0.13 & 2.16 & 1.64 & -0.40 & 0.40 \\
        mCSM & 3D & 0.37 & 1.96 & 1.49 & 0.36 & 1.54 & 1.13 & 0.22 & 2.30 & 1.86 & -0.05 & 0.85 \\
        Dynamut2 & 3D & 0.36 & 1.90 & 1.42 & 0.34 & 1.58 & 1.15 & 0.17 & 2.16 & 1.69 & 0.03 & 0.64 \\
        I-Mutant3.0 & 3D & 0.32 & 1.96 & 1.49 & 0.36 & 1.52 & 1.12 & 0.15 & 2.32 & 1.87 & -0.06 & 0.81 \\
        Rosetta & 3D & 0.47 & 2.69 & 2.05 & 0.39 & 2.70 & 2.08 & 0.40 & 2.68 & 2.02 & -0.72 & 0.61 \\
        FoldX & 3D & 0.31 & 2.39 & 1.53 & 0.22 & 2.30 & 1.56 & 0.22 & 2.48 & 1.50 & -0.20 & 0.34 \\
        SAAFEC-SEQ & SEQ & 0.26 & 2.02 & 1.54 & 0.36 & 1.54 & 1.13 & -0.01 & 2.40 & 1.94 & -0.03 & 0.83 \\
        MUpro & SEQ & 0.32 & 2.03 & 1.58 & 0.25 & 1.61 & 1.21 & 0.20 & 2.38 & 1.96 & -0.32 & 0.95 \\
        \hline
    \end{tabular}%
    }
\end{table*}

\begin{table*}[h]
    \centering
    \caption{Comparative benchmark of different sequence- and structure-based methods on the S461 independent test set of single-point variations.The data used to compute model performance, excluding JanusDDG, were taken from~\cite{reeves2024zero}.}
    \renewcommand{\arraystretch}{1.2}
    \begin{tabular}{lcccc}
        \hline
        \textbf{Model} & \textbf{Pearson} & \textbf{Spearman} & \textbf{RMSE} & \textbf{MAE} \\
        \hline
        \textbf{JanusDDG}  & \textbf{0.69} & \textbf{0.66} & \textbf{0.97} & \textbf{0.73} \\      
        Stability Oracle  & 0.61 & 0.63 & 1.19 & 0.89 \\
        CartDDG-D         & 0.60 & 0.61 & 3.59 & 2.93 \\
        PremPS            & 0.63 & 0.60 & 1.03 & 0.80 \\
        PopMusic          & 0.61 & 0.60 & 1.02 & 0.76 \\
        MAESTRO           & 0.63 & 0.60 & 1.04 & 0.79 \\
        INPS3D            & 0.61 & 0.59 & 1.02 & 0.76 \\
        DDGun3D           & 0.63 & 0.58 & 1.11 & 0.81 \\
        DUET              & 0.59 & 0.57 & 1.06 & 0.78 \\
        ACDC-NN           & 0.60 & 0.56 & 1.06 & 0.78 \\
        KORPMD            & 0.57 & 0.54 & 1.21 & 0.91 \\
        mCSM              & 0.53 & 0.51 & 1.07 & 0.81 \\
        SDM               & 0.56 & 0.51 & 1.33 & 1.02 \\
        ThermoNet         & 0.55 & 0.48 & 1.24 & 0.93 \\
        I-Mutant3.0       & 0.49 & 0.47 & 1.12 & 0.84 \\
        SAAFEC-Seq        & 0.49 & 0.47 & 1.12 & 0.84 \\
        MIF               & 0.45 & 0.46 & 4.37 & 3.14 \\
        Ankh              & 0.44 & 0.44 & 5.60 & 4.69 \\
        ESM2-650M         & 0.43 & 0.44 & 4.41 & 3.55 \\
        Dynamut           & 0.50 & 0.43 & 1.27 & 0.96 \\
        MPNN-20-00        & 0.40 & 0.43 & 2.36 & 1.88 \\
        ESM1v Mean        & 0.39 & 0.43 & 4.29 & 3.33 \\
        ESMIF Multimer    & 0.37 & 0.41 & 1.64 & 1.26 \\
        MIFST             & 0.37 & 0.38 & 5.02 & 3.95 \\
        MutComputeX       & 0.33 & 0.36 & 1.39 & 1.03 \\
        FoldX-D           & 0.30 & 0.39 & 1.91 & 1.26 \\
        Tranception       & 0.24 & 0.27 & 1.68 & 1.29 \\
        MSA Transformer Mean & 0.30 & 0.26 & 5.84 & 5.05 \\
        \hline
    \end{tabular}
    \label{tab:model_performance s461}
\end{table*}

\begin{table*}[h]
    \centering
    \caption{Performance comparison on s96. The data used to compute model performance, excluding JanusDDG, were taken from~\cite{montanucci2022ddgun}.}
    \begin{tabular}{lcccc}
        \hline
        \textbf{Model} & \textbf{Pearson} & \textbf{Spearman} & \textbf{RMSE} & \textbf{MAE} \\
        \hline
        JanusDDG  & \textbf{0.52} & \textbf{0.50} & \textbf{2.10} & \textbf{1.50} \\
        DDGun         & 0.48 & 0.44 & 2.14 & 1.59 \\
        DDGun3D       & \textbf{0.52} & 0.48 & \textbf{2.10} & 1.61 \\
        INPS-MD       & 0.43 & 0.37 & 2.21 & 1.67 \\
        Maestro       & 0.36 & 0.36 & 2.29 & 1.64 \\
        mCSM          & 0.31 & 0.33 & 2.33 & 1.72 \\
        FoldX         & 0.22 & 0.38 & 4.18 & 2.37 \\
        INPS          & 0.44 & 0.41 & 2.20 & 1.64 \\
        POPMUSIC      & 0.36 & 0.33 & 2.29 & 1.74 \\
        SDM           & 0.51 & 0.47 & 2.12 & 1.59 \\
        \hline
    \end{tabular}
    \label{tab:performance s96}
\end{table*}

\begin{table*}[h]
    \centering
    \caption{Performance comparison of different models on PTmul-NR test set. The models’ performance data, excluding JanusDDG, were taken from~\cite{savojardo2025ddgemb}.}
    \begin{tabular}{lcccc}
        \hline
        \textbf{Model} & \textbf{Pearson} & \textbf{RMSE} & \textbf{MAE} \\
        \hline
        DDG JanusDDG  & \textbf{0.61} & \textbf{2.06} & \textbf{1.57} & \\
        DDGemb         & 0.59 & 2.16 & 1.59 &  \\
        FoldX       & 0.36 & 5.51 & 3.66 & \\
        Maestro       & 0.28 & 2.55 & 1.88 &  \\
        DDGun         & 0.23 & 2.55 & 2.10 &  \\
        DDGun3D       & 0.17 & 2.57 & 2.08 \\
        \hline
    \end{tabular}
    \label{tab:performance ptmul_nr}
\end{table*}

\begin{table}[h]
    \centering
    \caption{Performance on Ssym. The data used to compute model performance, excluding JanusDDG, were taken from~\cite{reeves2024zero}.}
    \begin{tabular}{lcccccccccc}
        \toprule
        & \multicolumn{4}{c}{\textbf{Direct}} & \multicolumn{4}{c}{\textbf{Inverse}} & \multicolumn{2}{c}{\textbf{Antisymmetry}} \\
        \cmidrule(lr){2-5} \cmidrule(lr){6-9} \cmidrule(lr){10-11}
        \textbf{Dir} & \textbf{Pearson} & \textbf{Spearman} & \textbf{RMSE} & \textbf{MAE} & \textbf{Pearson} & \textbf{Spearman} & \textbf{RMSE} & \textbf{MAE} & \textbf{Pearson d-r} & \textbf{$<\delta>$} \\
        \midrule
        \textbf{JanusDDG}         & 0.84 &0.84 & 0.85  &0.55 & 0.84 &0.84 & 0.85  &0.55 & -1 & 0.00 \\
        KORPM                    & 0.56 & 0.58 & 1.30 & 0.94 & 0.49 & 0.51 & 1.40 & 1.00 & -0.88 & -0.11 \\
        mpnn\_20\_00              & 0.57 & 0.61 & 2.44 & 2.02 & 0.40 & 0.48 & 2.42 & 1.78 & -0.58 & -1.40 \\
        Cartddg                  & 0.66 & 0.69 & 3.32 & 2.66 & 0.45 & 0.43 & 3.56 & 2.63 & -0.41 & -3.13 \\
        ACDC-NN                  & 0.61 & 0.53 & 1.37 & 1.01 & 0.59 & 0.51 & 1.43 & 1.04 & -0.98 & -0.05 \\
        stability-oracle         & 0.65 & 0.68 & 1.38 & 0.95 & 0.42 & 0.42 & 1.77 & 1.26 & -0.57 & -0.50 \\
        mifst                    & 0.46 & 0.46 & 5.51 & 4.54 & 0.31 & 0.31 & 4.11 & 3.21 & -0.74 & -2.10 \\
        msa\_transformer\_mean    & 0.35 & 0.32 & 5.41 & 4.51 & 0.35 & 0.32 & 5.41 & 4.51 & -1.00 &  0.00 \\
        mif                      & 0.56 & 0.54 & 4.70 & 3.70 & 0.35 & 0.37 & 3.76 & 2.77 & -0.44 & -3.46 \\
        esm2\_650M                & 0.27 & 0.29 & 5.73 & 4.87 & 0.27 & 0.29 & 5.73 & 4.87 & -1.00 &  0.00 \\
        ankh                     & 0.28 & 0.29 & 6.06 & 5.26 & 0.28 & 0.29 & 6.06 & 5.26 & -1.00 &  0.00 \\
        tranception              & 0.26 & 0.25 & 1.83 & 1.33 & 0.26 & 0.25 & 1.83 & 1.33 & -1.00 &  0.00 \\
        esmif\_multimer           & 0.54 & 0.49 & 1.81 & 1.32 & 0.15 & 0.24 & 1.85 & 1.34 & -0.17 & -0.03 \\
        DDGun3D                  & 0.57 & 0.46 & 1.40 & 1.00 & 0.54 & 0.45 & 1.43 & 1.03 & -0.99 & -0.04 \\
        FoldX                    & 0.56 & 0.66 & 1.93 & 1.17 & 0.37 & 0.39 & 2.16 & 1.49 & -0.25 & -1.12 \\
        Evo                      & 0.58 & 0.54 & 1.36 & 1.00 & 0.32 & 0.31 & 1.75 & 1.26 & -0.58 & -0.36 \\
        mutcomputex              & 0.43 & 0.46 & 1.50 & 1.04 & 0.16 & 0.22 & 1.95 & 1.40 & -0.19 & -0.70 \\
        INPS3D                   & 0.61 & 0.58 & 1.24 & 0.89 & 0.29 & 0.15 & 1.94 & 1.45 & -0.51 & -1.02 \\
        esm1v\_mean               & 0.10 & 0.15 & 3.66 & 2.40 & 0.10 & 0.15 & 3.66 & 2.40 & -1.00 &  0.00 \\
        ThermoNet                & 0.45 & 0.39 & 1.57 & 1.10 & 0.37 & 0.31 & 1.66 & 1.16 & -0.85 & -0.04 \\
        MAESTRO                  & 0.57 & 0.60 & 1.31 & 0.91 & 0.27 & 0.24 & 2.16 & 1.66 & -0.33 & -1.25 \\
        DUET                     & 0.63 & 0.62 & 1.22 & 0.87 & 0.17 & 0.12 & 2.30 & 1.76 & -0.30 & -1.49 \\
        I-Mutant3.0              & 0.64 & 0.67 & 1.21 & 0.78 & -0.04 & -0.06 & 2.32 & 1.76 &  0.00 & -1.37 \\
        MUpro                    & 0.79 & 0.77 & 0.94 & 0.53 & 0.07 & 0.04 & 2.51 & 2.03 & -0.02 & -1.93 \\
        mCSM                     & 0.61 & 0.57 & 1.23 & 0.91 & 0.14 & 0.07 & 2.43 & 1.93 & -0.26 & -1.82 \\
        SDM                      & 0.50 & 0.50 & 1.57 & 1.22 & 0.17 & 0.14 & 2.34 & 1.80 & -0.43 & -1.09 \\
        Dynamut                  & 0.56 & 0.50 & 1.46 & 1.04 & 0.35 & 0.35 & 1.75 & 1.26 & -0.57 & -0.25 \\
        \bottomrule
    \end{tabular}
    \label{tab:ssym}
\end{table}

% \subsection{Performance on other Datasets}
% \label{sec:Performance on other Datasets}
% In this section, we present tables summarizing the performance of JanusDDG on additional datasets from the literature. However, a potential issue that may arise is the similarity between proteins in the training and test sets, which could influence the results. This similarity might lead to overestimating the model’s performance, making it essential to carefully evaluate the independence of the test datasets from the training data.

\begin{table*}[h]
    \centering
    \caption{Performance comparison of different models on on K2369. The models’ performance data, excluding JanusDDG, were taken from~\cite{reeves2024zero}.}
    \begin{tabular}{llcccccc}
        \toprule
        \textbf{Model Type} & \textbf{Model} & \textbf{MSE} & \textbf{Accuracy} & $\rho$ & $w\rho$ & \textbf{NDCG} & \textbf{wNDCG} \\
        \midrule

        unknown & $\Delta\Delta G_u$ label & $0.00 \pm 0.00$ & 1 & 1 & 1 & 1 & 1 \\
        sequence & \textbf{JanusDDG} & \textbf{1.14} & \textbf{0.72} & \textbf{0.70} & \textbf{0.56} & \textbf{0.87} & \textbf{0.83}\\
        ensemble & Ensemble 6 Feats* & $1.52 \pm 0.36$ & 0.73 & 0.66 & 0.5 & 0.81 & 0.75 \\
        ensemble & Ensemble 5 Feats* & $1.53 \pm 0.36$ & 0.73 & 0.65 & 0.51 & 0.82 & 0.74 \\
        ensemble & Ensemble 7 Feats* & $1.53 \pm 0.36$ & 0.73 & 0.66 & 0.5 & 0.81 & 0.74 \\
        ensemble & Ensemble 4 Feats* & $1.58 \pm 0.38$ & 0.72 & 0.65 & 0.51 & 0.82 & 0.75 \\
        transfer & Stability Oracle & $1.61 \pm 0.17$ & 0.7 & 0.59 & 0.48 & 0.75 & 0.76 \\
        ensemble & Ensemble 3 Feats* & $1.70 \pm 0.41$ & 0.72 & 0.59 & 0.45 & 0.81 & 0.74 \\
        potential & KORPM* & $1.72 \pm 0.35$ & 0.71 & 0.55 & 0.44 & 0.78 & 0.74 \\
        ensemble & Ensemble 2 Feats* & $1.96 \pm 0.49$ & 0.69 & 0.51 & 0.39 & 0.8 & 0.71 \\
        struc. PSLM & ESM-IF & $2.95 \pm 0.89$ & 0.65 & 0.4 & 0.41 & 0.76 & 0.71 \\
        seq. PSLM & Tranception (reduced) & $3.03 \pm 0.89$ & 0.6 & 0.31 & 0.24 & 0.71 & 0.68 \\
        seq. PSLM & Tranception & $3.03 \pm 0.89$ & 0.61 & 0.32 & 0.24 & 0.71 & 0.69 \\
        unknown & Gaussian Noise & $3.59 \pm 0.61$ & 0.53 & -0.02 & -0.02 & 0.65 & 0.59 \\
        struc. PSLM & ProteinMPNN 0.3 & $7.38 \pm 1.42$ & 0.66 & 0.47 & 0.42 & 0.81 & 0.73 \\
        struc. PSLM & ProteinMPNN 0.2 & $7.86 \pm 1.54$ & 0.67 & 0.47 & 0.42 & 0.78 & 0.73 \\
        struc. PSLM & ProteinMPNN 0.1 & $8.44 \pm 1.49$ & 0.66 & 0.47 & 0.4 & 0.79 & 0.73 \\
        seq. PSLM & ESM-2 150M & $22.0 \pm 5.61$ & 0.62 & 0.24 & 0.27 & 0.76 & 0.69 \\
        seq. PSLM & ESM-1V mean & $26.9 \pm 3.87$ & 0.63 & 0.26 & 0.28 & 0.76 & 0.68 \\
        seq. PSLM & ESM-2 650M & $29.4 \pm 4.54$ & 0.63 & 0.32 & 0.3 & 0.75 & 0.7 \\
        struc. PSLM & MIF & $30.7 \pm 6.37$ & 0.65 & 0.46 & 0.42 & 0.77 & 0.7 \\
        biophysical & Rosetta CartDDG & $32.4 \pm 3.52$ & 0.7 & 0.61 & 0.45 & 0.8 & 0.73 \\
        seq. PSLM & MSA-T mean & $32.7 \pm 5.98$ & 0.63 & 0.36 & 0.27 & 0.73 & 0.69 \\
        struc. PSLM & MIF-ST & $34.6 \pm 3.23$ & 0.64 & 0.45 & 0.38 & 0.77 & 0.71 \\
        seq. PSLM & Ankh & $37.7 \pm 3.53$ & 0.63 & 0.36 & 0.25 & 0.72 & 0.68 \\
        seq. PSLM & ESM-2 3B & $39.7 \pm 4.22$ & 0.62 & 0.32 & 0.31 & 0.71 & 0.69 \\
        seq. PSLM & ESM-2 15B & $46.0 \pm 3.54$ & 0.62 & 0.36 & 0.28 & 0.73 & 0.68 \\
        struc. PSLM & Rosetta/ProtMPNN & $66.1 \pm 7.09$ & 0.69 & 0.65 & 0.53 & 0.83 & 0.75 \\
        \bottomrule
    \end{tabular}
    \label{tab:model_comparison}
\end{table*}

\begin{table*}[h]
    \centering
    \caption{Performance on Q3421 with Alternative Choices for Statistics. The models’ performance data, excluding JanusDDG, were taken from~\cite{reeves2024zero}.}
    \begin{tabular}{llcccccc}
        \toprule
        \textbf{Model Type} & \textbf{Model} & \textbf{MSE} & \textbf{Accuracy} & \textbf{Spearman’s $\rho$} & \textbf{w$\rho$} & \textbf{NDCG} & \textbf{wNDCG} \\
        \midrule

        unknown & $\Delta\Delta$Gu label & 0.00 ± 0.00 & 1 & 1 & 1 & 1 \\ 
        struc. & \textbf{JanusDDG} & \textbf{2.10 ± } & \textbf{0.86} & \textbf{0.78} & \textbf{0.63} & \textbf{0.72} & \textbf{0.80} \\
        transfer & Stability Oracle & 2.98 ± 0.48 & 0.77 & 0.58 & 0.46 & 0.61 & 0.71 \\ 
        ensemble & Ensemble 5 Feats & 3.09 ± 0.42 & 0.75 & 0.59 & 0.48 & 0.6 & 0.68 \\ 
        ensemble & Ensemble 6 Feats & 3.09 ± 0.43 & 0.75 & 0.59 & 0.48 & 0.6 & 0.68 \\ 
        ensemble & Ensemble 7 Feats & 3.10 ± 0.42 & 0.75 & 0.59 & 0.48 & 0.6 & 0.69 \\ 
        ensemble & Ensemble 4 Feats & 3.20 ± 0.43 & 0.75 & 0.57 & 0.48 & 0.61 & 0.68 \\ 
        ensemble & Ensemble 3 Feats & 3.43 ± 0.45 & 0.72 & 0.5 & 0.4 & 0.59 & 0.66 \\ 
        potential & KORPM & 3.54 ± 0.48 & 0.73 & 0.47 & 0.34 & 0.59 & 0.66 \\ 
        ensemble & Ensemble 2 Feats & 3.68 ± 0.45 & 0.73 & 0.43 & 0.34 & 0.59 & 0.64 \\ 
        struc. & PSLM MutComputeX & 4.04 ± 0.44 & 0.78 & 0.36 & 0.28 & 0.56 & 0.64 \\ 
        unknown & Gaussian Noise & 4.96 ± 0.39 & 0.7 & 0 & 0 & 0.51 & 0.56 \\ 
        struc. & PSLM ESM-IF & 5.04 ± 0.49 & 0.77 & 0.44 & 0.41 & 0.6 & 0.66 \\ 
        seq. & PSLM Tranception (reduced) & 5.09 ± 0.50 & 0.77 & 0.25 & 0.22 & 0.53 & 0.58 \\ 
        seq. & PSLM Tranception & 5.09 ± 0.50 & 0.78 & 0.26 & 0.24 & 0.54 & 0.58 \\ 
        struc. & PSLM ProteinMPNN 0.3 & 7.26 ± 0.62 & 0.78 & 0.48 & 0.41 & 0.61 & 0.69 \\
        struc. & PSLM ProteinMPNN 0.2 & 7.57 ± 0.60 & 0.78 & 0.49 & 0.41 & 0.6 & 0.68 \\ 
        struc. & PSLM ProteinMPNN 0.1 & 8.33 ± 0.64 & 0.78 & 0.48 & 0.4 & 0.6 & 0.68 \\ 
        seq. & PSLM ESM-2 150M & 16.2 ± 2.14 & 0.72 & 0.22 & 0.24 & 0.57 & 0.62 \\ 
        struc. & PSLM MIF & 23.8 ± 1.89 & 0.77 & 0.47 & 0.4 & 0.6 & 0.68 \\
        seq. & PSLM ESM-1V mean & 24.2 ± 3.80 & 0.74 & 0.22 & 0.25 & 0.56 & 0.6 \\ 
        biophysical & Rosetta CartDDG & 26.7 ± 1.98 & 0.78 & 0.56 & 0.43 & 0.61 & 0.69 \\ 
        seq. & PSLM ESM-2 650M & 27.8 ± 2.64 & 0.76 & 0.29 & 0.3 & 0.57 & 0.63 \\
        struc. & PSLM MIF-ST & 34.8 ± 2.44 & 0.77 & 0.4 & 0.32 & 0.59 & 0.63 \\ 
        seq. & PSLM Ankh & 37.6 ± 2.74 & 0.77 & 0.31 & 0.28 & 0.55 & 0.62 \\ 
        seq. & PSLM MSA-T mean & 37.7 ± 3.01 & 0.77 & 0.28 & 0.24 & 0.56 & 0.61 \\ 
        seq. & PSLM ESM-2 3B & 42.8 ± 4.93 & 0.77 & 0.26 & 0.26 & 0.54 & 0.62 \\ 
        seq. & PSLM ESM-2 15B & 52.5 ± 5.34 & 0.77 & 0.25 & 0.25 & 0.55 & 0.61 \\ 
        struc. & PSLM Rosetta/ProtMPNN & 59.2 ± 3.80 & 0.8 & 0.62 & 0.49 & 0.62 & 0.71 \\ 
        \bottomrule
    \end{tabular}
    \label{tab:model_comparison}
\end{table*}

\begin{table*}[h]
    \centering
    \caption{Performance comparison of different methods on Ptmul-D. The performance of the ThermoMPNN-D model was taken from~\cite{Dieckhaus2025}.}
    \label{tab:metrics_multiple}
    \renewcommand{\arraystretch}{1.2}
    \begin{tabular}{l c c c}
        \hline
        \textbf{Method} & \textbf{PCC} & \textbf{SCC} &\textbf{RMSE}  \\
        \hline
        JanusDDG  & 0.55 & 0.55 & \textbf{1.91} \\
        ThermoMPNN-D   & \textbf{0.57} & \textbf{0.59} & 1.95 \\
        \hline
    \end{tabular}
\end{table*}

% \begin{table*}[h]
%     \centering
%     \caption{Performance comparison on validation set m28. The models’ performance data, excluding JanusDDG, were taken from~\cite{montanucci2022ddgun}.}
%     \begin{tabular}{lcccc}
%         \hline
%         \textbf{Model} & \textbf{Pearson} & \textbf{Spearman} & \textbf{RMSE} & \textbf{MAE} \\
%         \hline
%         DDGun         & 0.42 & 0.43 & 2.49 & 1.92 \\
%         DDGun3D       & 0.44 & 0.46 & 2.54 & 1.96 \\
%         Maestro       & 0.28 & 0.12 & 2.90 & 2.33 \\
%         FoldX         & 0.38 & 0.41 & 2.64 & 2.02 \\
%         JanusDDG  & \textbf{0.67} & \textbf{0.64} & \textbf{1.95} & \textbf{1.47} \\
%         \hline
%     \end{tabular}
%     \label{tab:performance}
% \end{table*}
\begin{table*}[h]
    \centering
    \caption{Performance comparison on s96. The models’ performance data, excluding JanusDDG, were taken from~\cite{montanucci2022ddgun}.}
    \begin{tabular}{lcc}
        \hline
        \textbf{Model} & \textbf{Pearson} & \textbf{RMSE} \\
        \hline
        JanusDDG  & \textbf{0.52} & \textbf{2.10} \\
        DDGun         & 0.48 & 2.14 \\
        DDGun3D       & \textbf{0.52} & \textbf{2.10} \\
        INPS-MD       & 0.43 & 2.21 \\
        Maestro       & 0.36 & 2.29 \\
        mCSM          & 0.31 & 2.33 \\
        FoldX         & 0.22 & 4.18 \\
        INPS          & 0.44 & 2.20 \\
        POPMUSIC      & 0.36 & 2.29 \\
        SDM           & 0.51 & 2.12 \\
        \hline
    \end{tabular}
    \label{tab:performance s96}
\end{table*}

\end{document}